\documentclass[%
longbibliography,
reprint,
superscriptaddress,
 amsmath,amssymb,
 aps,
 floatfix,
]{revtex4-1}

\usepackage{graphicx}
\usepackage{dcolumn}
\usepackage{lipsum}
\usepackage{bm}
\usepackage{xcolor}
\usepackage{braket} 
\usepackage[colorlinks,linkcolor=blue,citecolor=blue,urlcolor=blue]{hyperref}
\begin{document}


\title{The phase sensitivity of a fully quantum three-mode nonlinear interferometer}

\author{Jefferson Fl\'orez}
\email{jflor020@uottawa.ca}
\affiliation{Department of Physics and Centre for Research in Photonics, University of Ottawa, 25 Templeton Street, Ottawa, Ontario K1N 6N5, Canada}
\author{Enno Giese}
\altaffiliation[]{Current affiliation: Institut f{\"u}r Quantenphysik and Center for Integrated Quantum Science and Technology (IQ$^\mathrm{ST}$), Universit{\"a}t Ulm, Albert-Einstein-Allee 11, D-89069 Ulm, Germany}
\author{Davor Curic}
\affiliation{Department of Physics and Centre for Research in Photonics, University of Ottawa, 25 Templeton Street, Ottawa, Ontario K1N 6N5, Canada}
\author{Lambert Giner}
\affiliation{Department of Physics and Centre for Research in Photonics, University of Ottawa, 25 Templeton Street, Ottawa, Ontario K1N 6N5, Canada}
\author{Robert W. Boyd}
\affiliation{Department of Physics and Centre for Research in Photonics, University of Ottawa, 25 Templeton Street, Ottawa, Ontario K1N 6N5, Canada}
\affiliation{Institute of Optics, University of Rochester, Rochester, New York 14627, USA}
\author{Jeff S. Lundeen}
\affiliation{Department of Physics and Centre for Research in Photonics, University of Ottawa, 25 Templeton Street, Ottawa, Ontario K1N 6N5, Canada}




\date{\today}

\begin{abstract}
We study a nonlinear interferometer consisting of two consecutive parametric amplifiers, where all three optical fields (pump, signal and idler) are treated quantum mechanically, allowing for pump depletion and other quantum phenomena.
The interaction of all three fields in the final amplifier leads to an interference pattern from which we extract the phase uncertainty.
We find that the phase uncertainty oscillates around a saturation level that decreases as the mean number $N$ of input pump photons increases.
For optimal interaction strengths, we also find a phase uncertainty below the shot-noise level and obtain a Heisenberg scaling $1/N$. 
This is in contrast to the conventional treatment within the parametric approximation, where the Heisenberg scaling is observed as a function of the number of \emph{down-converted} photons inside the interferometer.
\end{abstract}

\maketitle


\section{Introduction}
\label{sec_Intro}

The advantage of non-classical light in interferometry is one of the major applications of quantum mechanics to metrology.
For example, phase measurements with Mach-Zehnder interferometers illuminated by a classical coherent light field are limited by shot noise.
Therefore, their phase uncertainty scales as $1/\sqrt{N}$, where $N$ is the mean number of photons input to the interferometer.
However, non-classical input states, such as squeezed light, provide a \emph{Heisenberg scaling} of the phase uncertainty with $1/N$ for the same mean number of input photons~\cite{Caves81}.
Squeezed states are generated by nonlinear optical processes, in particular by parametric amplification.
The idea of integrating nonlinear optical elements directly into the structure of an interferometer, and using amplifiers instead of beam splitters, led to a new class of devices called \emph{nonlinear interferometers} (NLIs)~\cite{Yurke86}.

NLIs can be characterized by the Lie group SU(1,1) and exhibit phase uncertainty below shot noise~\cite{Chekhova16}.
Because of this phase sensitivity, NLIs constitute a possible alternative in optical quantum metrology~\cite{Hudelist14}.
Beyond this application, they have been used for spectroscopy~\cite{Kalashnikov16} and imaging with entangled photons of different colors~\cite{BarretoLemos14}.
NLIs also serve to shape and generate bright radiation with quantum properties~\cite{Lemieux16}. The concept can be applied to hybrid atom-light systems to study nonlinear dynamics in entangled systems~\cite{Linnemann16} or perform magnetometry beyond the shot-noise level~\cite{Chen15}.

\begin{figure}[htbp]
\centering
\includegraphics[width=\linewidth]{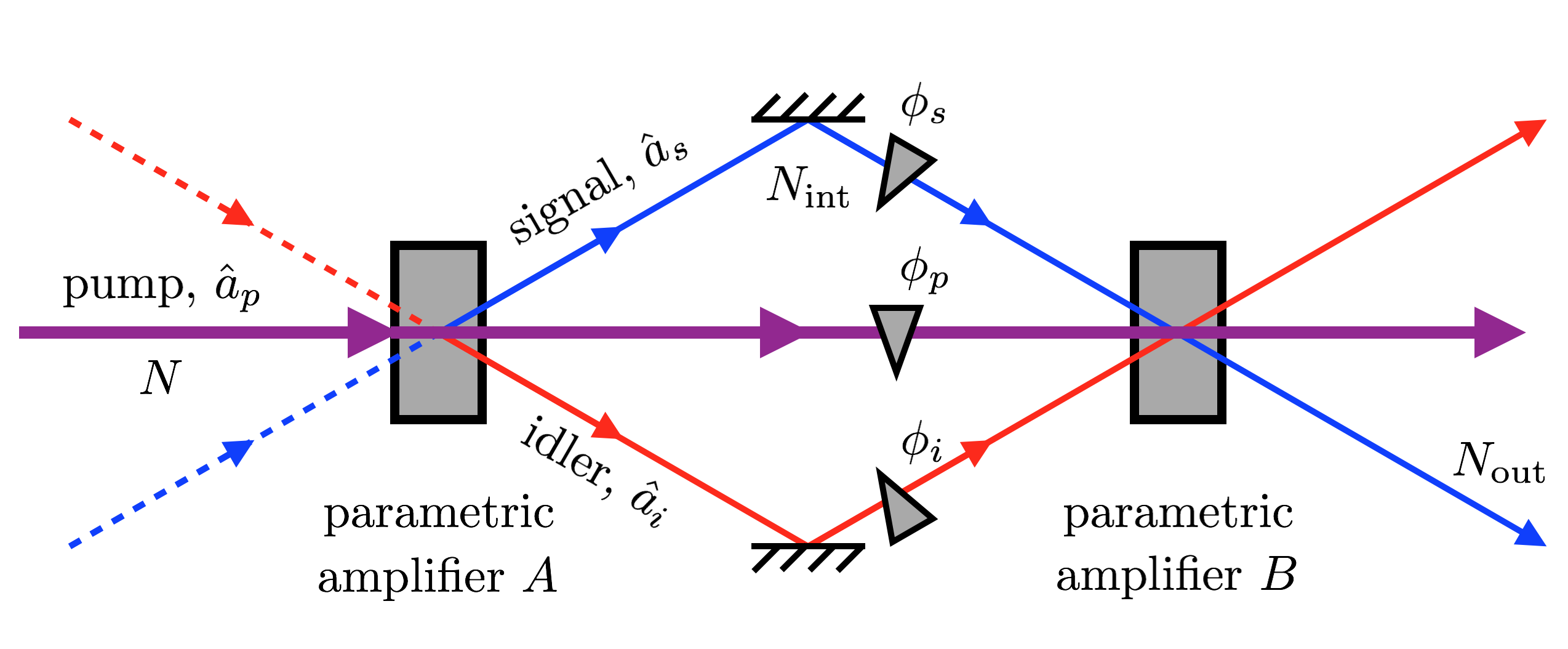} 
\caption{Schematic of a nonlinear interferometer.
A parametric amplifier $A$ populates signal and idler fields $\hat{a}_s$ and $\hat{a}_i$ (each with mean number $N_\mathrm{int}$ of internal photons) from a pump $\hat{a}_p$ (with photon mean number $N$).
All three fields acquire a phase $\phi_s$, $\phi_i$, and $\phi_p$, respectively, before entering a second parametric amplifier $B$.
The three output fields are detected and the mean number $N_\mathrm{out}$ of output signal (or idler) photons exhibits an interference pattern as a function of the phase difference between the three fields. The dashed lines entering amplifier $A$ signify that the signal and idler input fields are in the vacuum state.}
\label{f_NLI} 
\end{figure}

In this article, we focus on a particular type of NLI, consisting of two parametric amplifiers, $A$ and $B$, as shown in Fig.~\ref{f_NLI}.
Such a parametric amplifier usually consists of a medium with $\chi^{(2)}$ nonlinear optical properties (like beta barium borate crystal) pumped by a coherent field.
This device can be used to amplify an input signal field, in a second-order nonlinear optical process known as difference-frequency generation, or to generate two output fields by spontaneous down-conversion~\cite{Boyd08}.
We discuss here the case where only the pump ($p$) field contains photons at the input side of the NLI~\cite{Manceau17b,Linnemann16}, with a photon mean number $N$, even though different input fields can be used~\cite{Plick10,Li14,Sparaciari16}.

\

In a standard NLI, the signal ($s$) and idler ($i$) photons down-converted by the parametric amplifier $A$ are used as the input for amplifier $B$, as shown in Fig.~\ref{f_NLI}.
The pump field exiting amplifier $A$ is used as well to pump amplifier $B$ in most experimental realizations since the two amplifiers have to be pumped coherently.
The three fields (pump, signal and idler) then illuminate amplifier $B$ after acquiring a phase $\phi_j$ ($j=p,s,i$) upon propagation between amplifiers $A$ and $B$.
In amplifier $B$, a second parametric amplification process occurs, generating signal and idler output fields, each with a photon mean number $N_\mathrm{out}$, plus the pump.
In a linear interferometer, like a Mach-Zehnder setup, the interference can be explained through the indistinguishability of the two paths of the interferometer.
In an NLI, both amplifiers emit into the same modes and therefore it is impossible to distinguish whether it was amplifier $A$ or, rather, $B$ that created the signal and idler photons at the NLI output. This indistinguishability results in interference, whose interference pattern depends on the phase difference $\phi_p-\phi_s-\phi_i\equiv\phi$.

NLIs have attracted some attention~\cite{Chekhova16} due to the Heisenberg scaling of their phase uncertainty with the number $N_\mathrm{int}^{(\mathrm{PA})}$ of \emph{internal} signal (or idler) photons in the interferometer.
The superscript PA stands for parametric approximation, which will be explained below.
More precisely, the lowest phase uncertainty, which happens at the phase giving destructive interference, is~\cite{Yurke86}
\begin{equation}
\Delta \phi_\mathrm{PA}=\left[4 N_\mathrm{int}^\mathrm{(PA)}\left(1+N_\mathrm{int}^\mathrm{(PA)}\right) \right]^{-1/2}.
\label{e_PA}
\end{equation}
Consequently, Heisenberg scaling is reached when $N_\mathrm{int}^\mathrm{(PA)}\gg 1$.
In fact, a sub-shot noise sensitivity has been demonstrated experimentally~\cite{Linnemann16,Manceau17b}.

Equation~\eqref{e_PA} has been obtained within the \emph{parametric approximation} (PA), which assumes that the pump is an intense and undepleted classical field~\cite{Mollow67}.
That is, its quantum nature is neglected.
In this approximation, the number of signal (or idler) photons inside the NLI takes the form $N_\mathrm{int}^\mathrm{(PA)} = \sinh^2(\sqrt{N}\tau)$, where $N$ is the mean number of input pump photons and $\tau$ the nonlinear interaction strength (proportional to the second order electric susceptibility and the length of the nonlinear crystal).
Under the parametric approximation, the uncertainty $\Delta \phi_\mathrm{PA}$ scales as $1/\exp(2\sqrt{N}\tau )$ for strong gain, i.e. $\sqrt{N}\tau \gg 1$.
This suggests that $\Delta \phi_{\mathrm{PA}}$ appears to follow a \emph{super}-Heisenberg scaling with $N$.
However, this scaling implicitly violates energy conservation, since the number $N_\mathrm{int}^\mathrm{(PA)}$ of generated photons grows exponentially with $N$. 
This points to the possibility that pump depletion, and perhaps even the quantum features of the pump, play a crucial role for the sensitivity of an NLI.
In fact, experiments using even a small number of pump photons have shown that the quantum nature of the pump can be important in some cases~\cite{Hamel14,Ding17}.

In this article, we find the limit for the phase sensitivity of an NLI by taking into account the quantum nature of the the pump and its evolution during the amplification processes.
We show that quantum phenomena occurring in a single amplifier, like pump depletion, single mode squeezing, and entanglement between all three optical fields~\cite{Drobny93}, significantly contribute to the phase sensitivity.
Furthermore, we demonstrate that under certain conditions the phase uncertainty displays a Heisenberg scaling with the mean number $N$ of input pump photons.

We start this article by discussing the implemented numerical method in Sec.~\ref{sec_NLI}.
In Sec.~\ref{sec_Uncertainty}, we obtain the phase sensitivity for different input states for the pump, including coherent and Fock states, and compare it to the parametric approximation. 
We also show a Heisenberg scaling of the phase uncertainty with $N$ for optimized interaction strengths.
In Sec.~\ref{sec_Inside}, we take a closer look at the states inside the NLI that give the highest phase sensitivity, and discuss different possible reasons for this behavior.
To keep this article self-contained, we include Appendix \ref{app_CFI}, where the phase uncertainty is studied in terms of the classical Fisher information.

\section{The nonlinear interferometer}
\label{sec_NLI}
In this section, we formally investigate the interferometer shown in Fig.~\ref{f_NLI}, where two parametric amplifiers mix the pump, signal and idler fields.
These three fields are respectively associated with annihilation operators $\hat{a}_p$, $\hat{a}_s$, and $\hat{a}_i$, so that their interaction in each parametric amplifier is described through the trilinear Hamiltonian~\cite{Dicke54,Tavis68,Tucker69,Walls70,Bonifacio70}
\begin{equation}
\hat{H}(\theta)=\kappa e^{i\theta} \hat{a}_p \hat{a}_s^\dagger \hat{a}_i^\dagger+\kappa e^{-i\theta} \hat{a}_p^\dagger \hat{a}_s \hat{a}_i.
\label{e_tri}
\end{equation}
Here, we have introduced a generic optical phase $\theta_j$ on each of the three fields ($j=p,s,i$), and see that only the phase difference $\theta_p-\theta_s-\theta_i\equiv\theta$ appears explicitly. 
Note also that $\kappa$ denotes the (real) coupling strength, proportional to the nonlinear susceptibility of the nonlinear crystal, and that we chose $\hbar=1$.

In the parametric approximation, $\hat{a}_p$ and $\hat{a}_p^\dagger$ are respectively replaced by $\alpha$ and $\alpha^*$ in Eq.~\eqref{e_tri}, where $\alpha$ is a complex number such that $|\alpha|^2=N$ is the normalized pump intensity.
In this case, the resulting Hamiltonian can be solved analytically, leading to the unbounded exponential increase of the internal number $N_\mathrm{int}^\mathrm{(PA)}$ of signal (and idler) photons in the interferometer, as discussed in Sec.~\ref{sec_Intro}. To study the effect of both pump depletion and the quantum features of the pump, the parametric approximation cannot be made anymore.
However, there is no analytic solution for the states produced by the trilinear Hamiltonian in Eq.~\eqref{e_tri}.
Therefore, we follow Refs.~\cite{Walls70,Drobny92} and solve the Schr\"odinger equation $i\partial\ket{\psi(t)}/\partial t= \hat{H} \ket{\psi(t)}$ through a numerical diagonalization of the trilinear Hamiltonian in a basis composed by Fock states of the form
\begin{equation}
\ket{\nu}^{(N)} \equiv \ket{N-\nu}_{p} \ket{\nu}_{s}\ket{\nu}_{i},\quad \nu=0,1,\dots,N.
\label{e_basis}
\end{equation}
Here, $\nu$ is the number of annihilated pump photons and, at the same time, the number of photons generated in the signal and idler mode if the pump was initially in the Fock state $\ket{N}_p$ and the other fields in their vacuum state.
For these initial photon numbers, the state after an interaction time $t$ (proportional to the nonlinear crystal length) in the amplifier can be decomposed in the basis $\{\ket{\nu}^{(N)}\}$ as
\begin{equation}
\ket{\psi(t)}=\sum_{\nu=0}^{N}c_\nu(t)\ket{\nu}^{(N)},
\label{e_decomposition}
\end{equation}
where we introduced time-dependent complex coefficients $c_\nu(t)$.

Using the relation $\hat{a}_j \ket{n}_j = \sqrt{n} \ket{n-1}_j$ and the decomposition from Eq.~\eqref{e_decomposition}, we find from the Schr\"odinger equation a system of coupled differential equations for the coefficients,
\begin{equation}
i\dot{c}_\nu(t)=\kappa \left[m_{\nu-1} c_{\nu-1}(t)+m_\nu^* c_{\nu+1}(t) \right].
\label{e_recurrence}
\end{equation}
Here, we define the phase-dependent quantity $m_\nu= (\nu+1)\sqrt{N-\nu}\exp(i\theta)$, which vanishes for $\nu=N$.
Hence, the three-term recurrence relation terminates and we do not need to introduce an additional truncation.
We define the $(N+1)\times(N+1)$-matrix $M(\theta)$ with $M_{\nu,\mu}= m_{\nu-1} \delta_{\mu,\nu-1} + m_{\nu}^* \delta_{\mu,\nu+1}$ matrix elements, as well as a vector $\mathbf{c}^\mathrm{T}= (c_0,c_1,...,c_N)^\mathrm{T}$ describing the quantum state. We find the solution of Eq.~\eqref{e_recurrence} by numerically diagonalizing the Hermitian
coupling matrix $M(\theta)$. The resulting solution is given by
$\mathbf{c}(t)=\exp\left[ - i \tau M(\theta)\right] \mathbf{c}(0)$.
Here, we have introduced the dimensionless interaction strength $\tau = \kappa t $.
In the parametric approximation, a gain can be defined as $g= \sqrt{N} \tau $.

We now calculate the evolution of the three fields through the NLI. First, we obtain the output of amplifier $A$
\begin{equation}
\mathbf{c}^{(A)}(\tau)=\exp\left[ - i \tau M(0)\right] \mathbf{c}_\mathrm{in},
\label{e_cA}
\end{equation}
where $\mathbf{c}_\mathrm{in}$ is the input of the NLI.
The output of the interferometer after amplifier $B$ is
\begin{equation}
\mathbf{c}^{(B)}(\tau)=\exp\left[ - i \tau M(\phi)\right] \mathbf{c}^{(A)}.
\label{e_cB}
\end{equation}
Without loss of generality, we have choosen the phase $\theta=0$ for amplifier $A$ as the reference phase, and set $\theta =\phi$ in for amplifier $B$.
Note further that we have assumed an equal interaction strength $\tau$ in both amplifiers.
However, our treatment could be generalized to a gain-unbalanced situation in analogy to~\cite{Manceau17a,Manceau17b,Giese17}, which discuss the benefits of different coupling strengths within the parametric approximation in lossy NLIs~\cite{Marino12}.
Since we only investigate a lossless NLI, there would be no benefit from unbalancing the gain parameters. Therefore, we focus on the balanced configuration in this article.

We discuss two different pump input states in this article: a Fock state $\ket{N}_p$ and a coherent state $\ket{\alpha}_p$.
We shall use the symbol $N$ to denote the mean number of input pump photons in both cases, Fock states and coherent states, with $N\equiv|\alpha|^2$ in the latter case.
The signal and idler fields are always initially in a vacuum state.
Hence, note that there are always perfect correlations between the number of photons in the idler and signal fields.
For a Fock input state with $N$ pump photons, we have $\ket{\psi_\mathrm{in}} = \ket{0}^{(N)}$, which means $\mathbf{c}_\mathrm{in}^\mathrm{T}=(1,0,\dots,0)^\mathrm{T}$, and see that our state can be decomposed in the $\{|\nu\rangle^{(N)}\}$ basis at any time.

For a coherent input state, given by
\begin{equation}
\ket{\psi_\mathrm{in}}= \operatorname{e}^{-|\alpha|^2/2}\sum_{n=0}^\infty \frac{\alpha^n}{\sqrt{n!}}\ket{0}^{(n)},
\label{e_Coherent}
\end{equation}
we use the linearity of the Schr\"odinger equation to propagate each state $\ket{0}^{(n)}$ individually, using the method described above.
We truncate all states in Eq.~\eqref{e_Coherent} whose population is smaller than the population of the state $\ket{0}^{(N)}$ times $10^{-5}$, as a balance between numerical accuracy and computational time.

\section{Phase uncertainty}
\label{sec_Uncertainty}
With the treatment from Sec.~\ref{sec_NLI} we numerically find the quantum state of the pump, signal and idler fields inside the interferometer.
From this state we calculate the mean number $N_\mathrm{int}$ of internal signal (or idler) photons.
We use this number later to investigate whether the NLI phase uncertainty is in fact given by Eq.~\eqref{e_PA} if we set $N_\mathrm{int}^\text{(PA)}=N_\mathrm{int}$.

\begin{figure}[htbp]
\centering
\includegraphics[width=\linewidth]{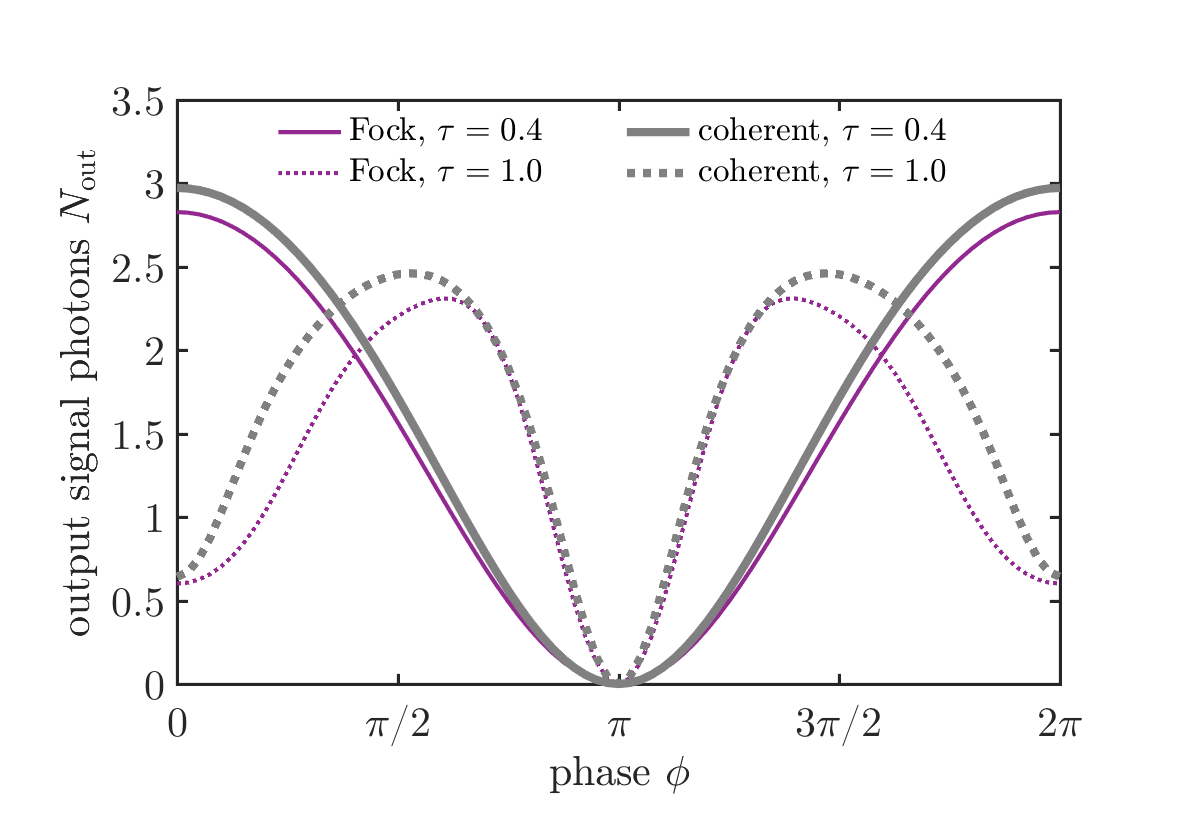} 
\caption{Interference patterns at the output of the nonlinear interferometer. 
The input pump field is in either a Fock or a coherent state, both with a mean photon number $N=5$.
For all the different interaction strengths $\tau$, the output number $N_\text{out}$ of signal (or idler) photons identically vanishes at $\phi=\pi$.}
\label{f_pattern} 
\end{figure}

From Eq.~\eqref{e_cB} we can obtain the mean number $N_\mathrm{out}$ of signal (or idler) photons at the output of the NLI. 
In Fig.~\ref{f_pattern}, we show the resulting interference pattern. 
That is, we plot $N_\mathrm{out}$ as a function of the phase $\phi$ in the interferometer for different input pump states and interaction strengths $\tau$.
Since $N_\mathrm{out}$ is phase sensitive, it can be used to estimate the phase of the interferometer by inverting the relevant curve in Fig.~\ref{f_pattern}.
Together with the error propagation formula, it can then be used to find the NLI phase uncertainty \cite{Gerry04},
\begin{equation}
\Delta \phi=\left.\frac{\sqrt{\mathrm{Var}(N_\mathrm{out})}}{\left|\partial N_\mathrm{out}/\partial \phi\right|}\right\rvert_{\phi=\pi+\delta}.
\label{e_EP}
\end{equation}
Here, $\mathrm{Var}(N_\mathrm{out})$ denotes the variance of the number of signal photons at the output of the NLI. 
For completeness, we present in Appendix~\ref{app_CFI} the phase uncertainty estimated from the classical Fisher information, but find qualitatively the same results as the ones obtained by means of Eq.~\eqref{e_EP}.

\begin{figure*}[t!]
\centering
\includegraphics[width=0.0437\linewidth]{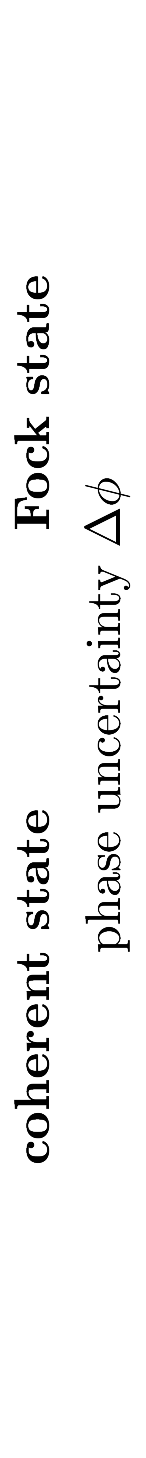}
\includegraphics[width=0.312\linewidth]{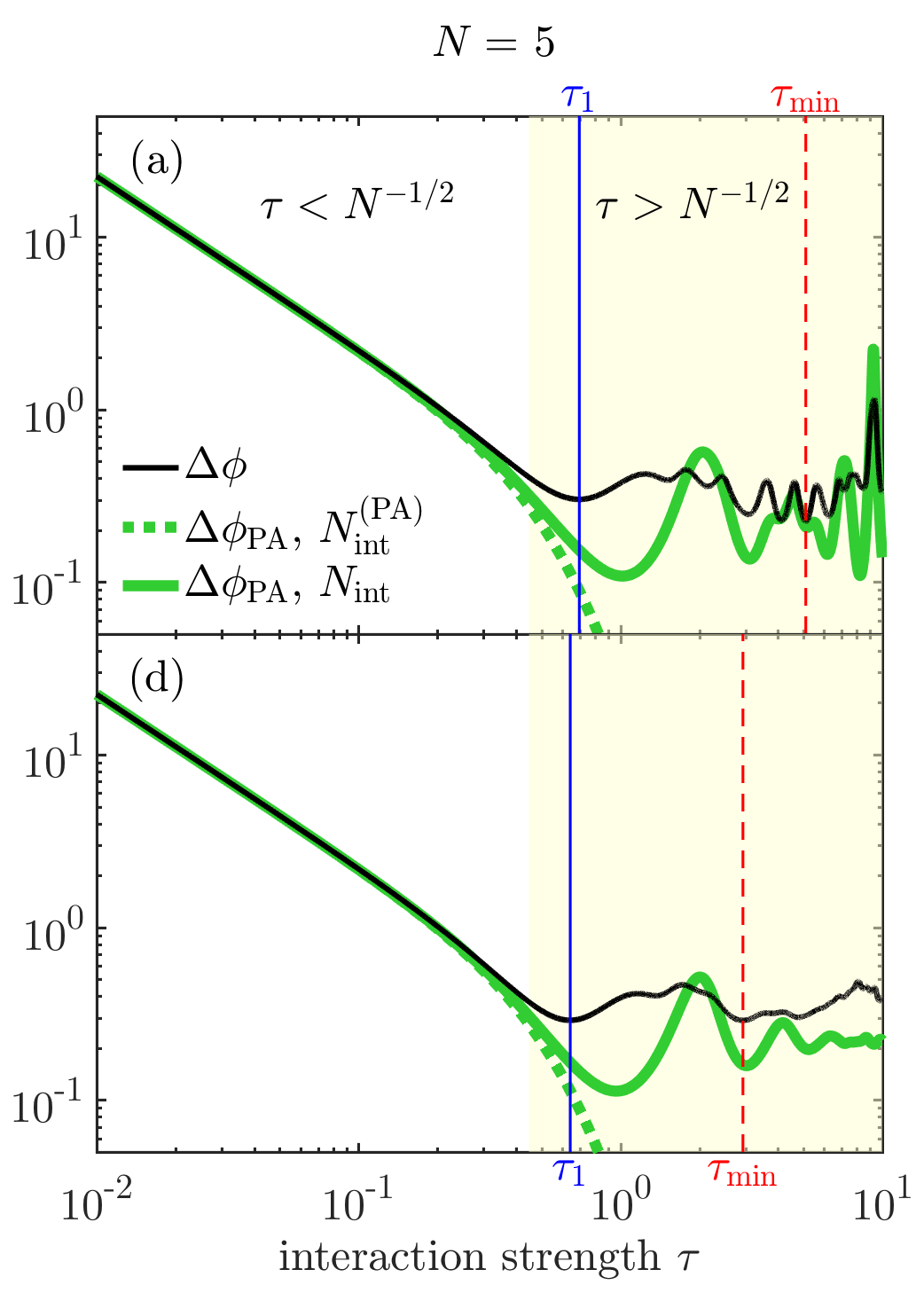}
\includegraphics[width=0.312\linewidth]{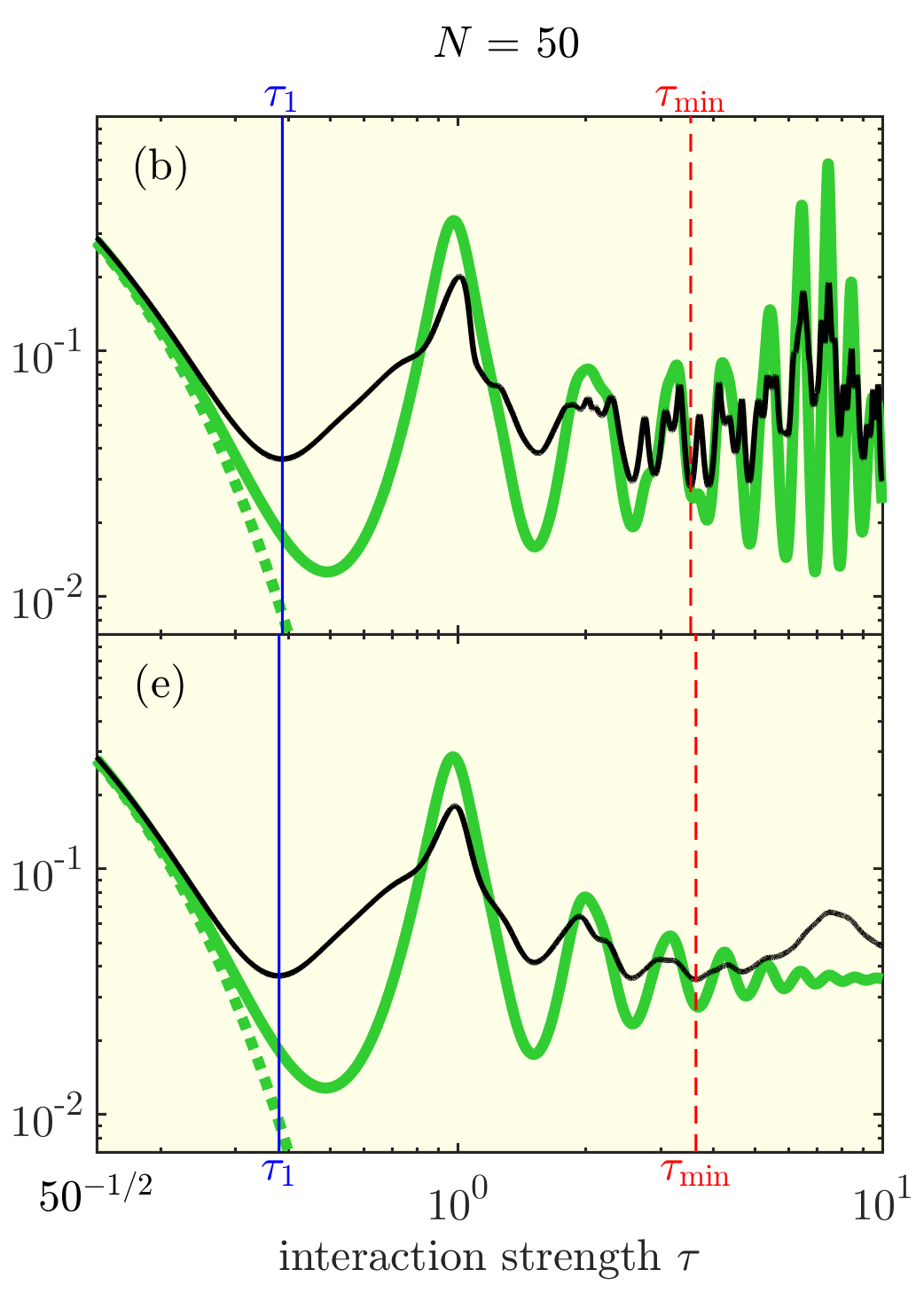}
\includegraphics[width=0.312\linewidth]{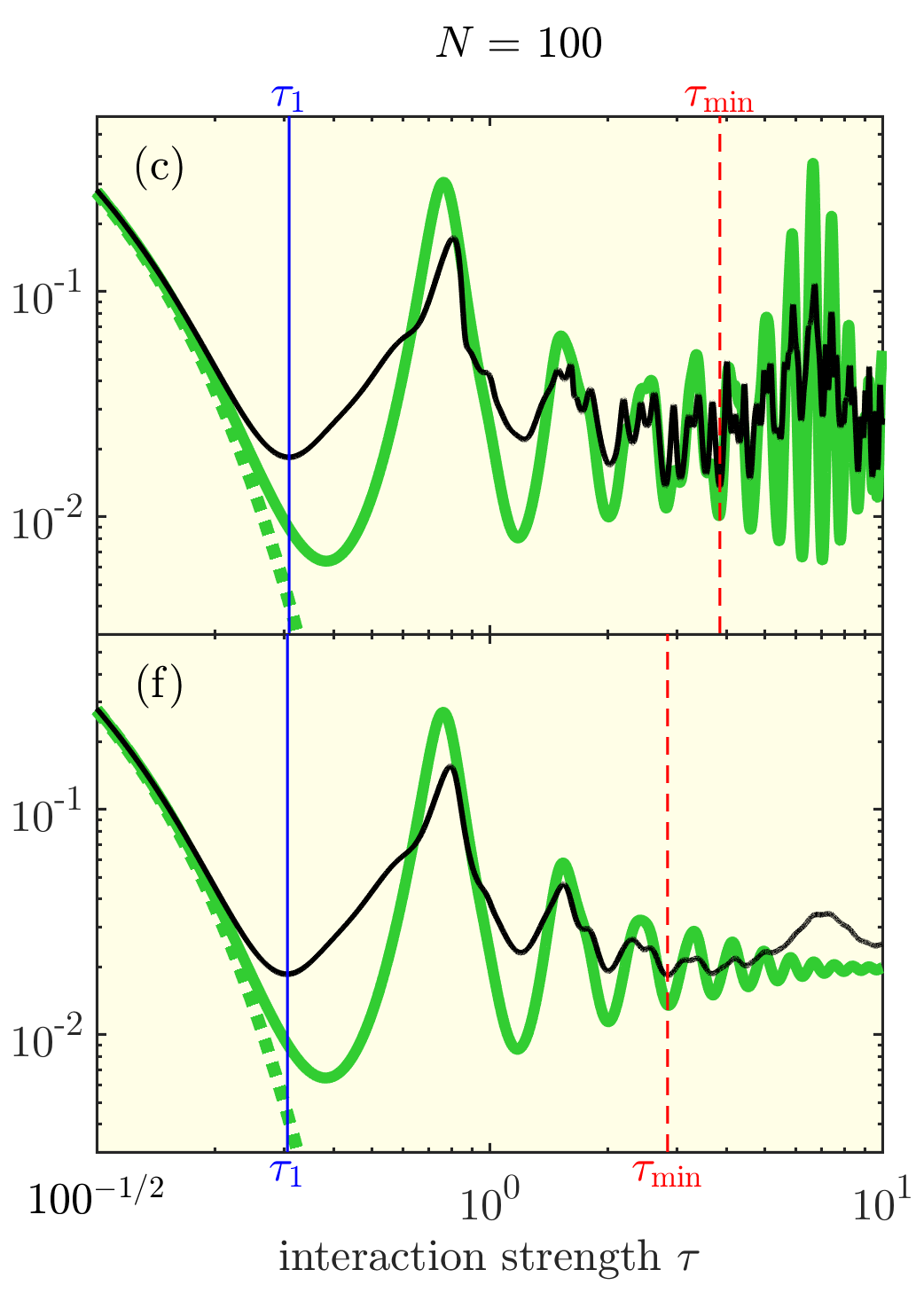}
\caption{Phase uncertainty $\Delta \phi$ of the nonlinear interferometer for different interaction strengths $\tau$ and mean number $N$ of input pump photons.
Panels (a-c) correspond to an input pump in a Fock state $|N\rangle_p$, whereas panels (d-f) correspond to an input pump in a coherent state $|\alpha\rangle_p$, with $N\equiv|\alpha|^2$.
In each panel, we compare the phase uncertainty $\Delta \phi$ to the parametric approximation result $\Delta\phi_\mathrm{PA}$ from Eq.~\eqref{e_PA}.
To illustrate the effect of pump depletion, we replace $N_\mathrm{int}^\text{(PA)}$ in the parametric approximation by the numerically obtained number $N_\mathrm{int}$ of internal signal (or idler) photons in the interferometer. 
The vertical lines in each panel indicate the interaction strengths $\tau_1$ and $\tau_\mathrm{min}$ at which the first and lowest $\Delta\phi$ minimum are observed, respectively. The yellow shadow area defines the high-gain regime, $\tau>N^{-1/2}$. In particular, panels (b,c,e,f) only display the phase uncertainty results in this regime.}
\label{f_Unc} 
\end{figure*}

The phase uncertainty $\Delta\phi$ in Eq.~\eqref{e_EP} can be calculated for any $\phi$ value. 
However, from the interference pattern in Fig.~\ref{f_pattern}, we observe that the output signal and idler fields are in the vacuum state at $\phi=\pi$ for all input pump states and interaction strengths.
The interference patterns for all other pump intensities $N$ are qualitatively the same, exhibiting in particular perfect destructive interference at $\phi=\pi$.
For this phase, the parametric amplifier $B$ reverses the unitary transformation performed by amplifier $A$, returning the input state, which was the vacuum state of the signal and idler fields.
Since the vacuum state is a photon number eigenstate,  $\mathrm{Var}(N_\mathrm{out})=0$. 
Given Eq.~\eqref{e_EP}, in turn this suggests that the phase uncertainty $\Delta\phi$ will be low at $\phi=\pi$. 
However, $N_\text{out}$ also exhibits a minimum at $\phi=\pi$ according to Fig.~\ref{f_pattern}, leading to a vanishing derivative in Eq.~\eqref{e_EP}. 
To properly obtain the derivative, we calculate $\Delta\phi$ at $\phi=\pi+\delta$, with $\delta\to 0$. 
Thus, the derivative in Eq.~\eqref{e_EP} reduces to the asymmetric difference quotient $\left.\partial N_\mathrm{out}/\partial \phi\right \rvert_{\pi+\delta}= \left.N_\text{out}\right \rvert_{\pi+2\delta}/(2\delta)$, where we have used $\left.N_\text{out}\right \rvert_{\pi}=0$.
We choose $\delta=\pi\times10^{-9}/2$ as a compromise between $N_\mathrm{out}$ be evaluated as close as possible to $\pi$, and keeping enough precision digits in our calculations, which is limited to 16 digits.

The results of our simulations are shown in Fig.~\ref{f_Unc} for a pump in either a Fock (top) or a coherent (bottom) state, and three different mean number of input photons, $N=5$, 50, and 100 (from left to right).
From Fig.~\ref{f_Unc}(a,d), we see that in the low-gain regime, $\tau<N^{-1/2}$, the uncertainty $\Delta \phi$ (black thin line) coincides perfectly with the parametric approximation uncertainty $\Delta \phi_\text{PA}$ (green thick dotted line) and displays an exponential scaling.
However, in this regime $N_\mathrm{int}^\text{(PA)}=\sinh^2 (\sqrt{N}\tau)\lesssim1$ and therefore there is no benefit from the Heisenberg scaling.
To have larger photon numbers inside the NLI and to benefit from the Heisenberg scaling, we need to enter the high-gain regime, i.e. $\tau > N^{-1/2}$ (yellow shaded area).
However, in this regime $\Delta \phi$ and $\Delta \phi_\text{PA}$ deviate significantly as the parametric approximation breaks down. 
In particular, $\Delta \phi$ begins oscillating in a non-periodic manner around a saturation level that decreases as $N$ increases. 
For a coherent state pump, these oscillations are somewhat smoother.
To further appreciate the oscillatory $\Delta\phi$ behaviour in the high-gain regime, we focus our analysis on interaction strengths $\tau \geq N^{-1/2}$ in Fig.~\ref{f_Unc}(b,c,e,f).

In particular, we discuss whether the sensitivity of the NLI in this regime is dictated by the number $N_\mathrm{int}$ of internal signal photons, which we calculate numerically. 
Even in classical nonlinear optics, one expects the pump to deplete with increasing interaction strength.
Therefore, the exponential growth of the number of generated photons will fall off. 
We examine whether this fall off fully explains the saturation and oscillation in the phase uncertainty shown in Fig.~\ref{f_Unc}.
As shown in Ref.~\cite{Walls70}, indeed $N_\mathrm{int}$ oscillates, which reflects a back and forth energy exchange between the pump and signal (and idler) fields after amplifier $A$.
Therefore, by simply replacing $N_\mathrm{int}^\text{(PA)}$ by $N_\mathrm{int}$ in Eq.~\eqref{e_PA} one predicts oscillatory behavior of the phase uncertainty $\Delta \phi_\text{PA}$ (green thick solid line in Fig.~\ref{f_Unc}).
However, while this \textit{ad hoc} substitution predicts a behavior similar to the exact phase uncertainty $\Delta \phi$, it does not describe all its features.
In particular, $\Delta \phi$ contains finer oscillations and does not go as low or high as $\Delta \phi_\text{PA}$ calculated from $N_\mathrm{int}$.
Hence, the phase sensitivity is not solely determined by the number of signal (or idler) photons inside the interferometer, and thus, not solely by pump depletion.
This suggests that the features found in $\Delta\phi$ are instead due to a combination of causes.
These could include the depletion of the pump, the quantum features of the pump (like single-mode squeezing), and entanglement between all three fields.

\vspace{-0.2cm}
\begin{center}
\begin{figure}[htbp]
\centering
\includegraphics[width=\linewidth]{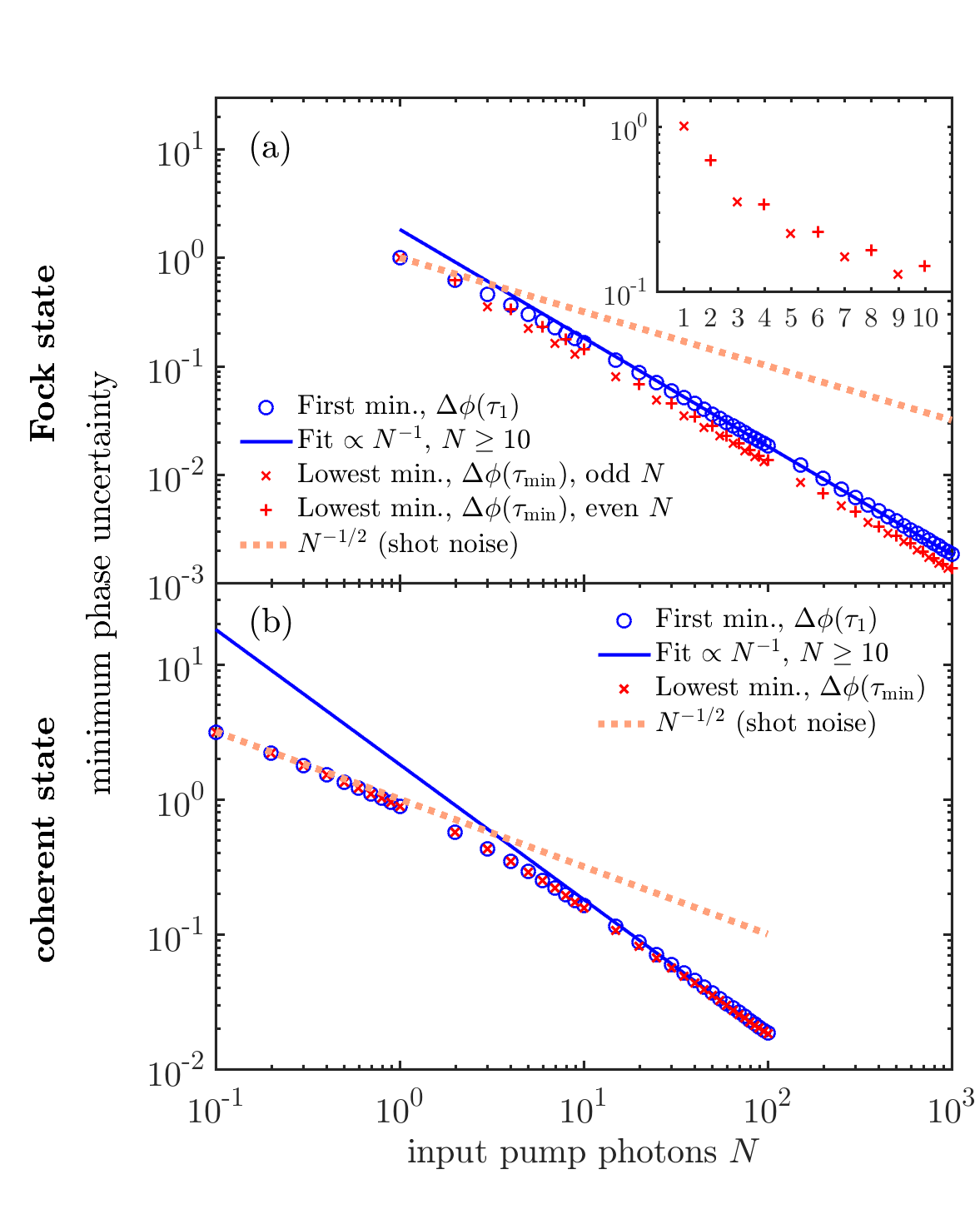}
\caption{First and lowest minimum of the phase uncertainty $\Delta\phi$. 
The input pump is in either a Fock (a) or a coherent (b) state. 
We show a Heisenberg scaling by fitting the first minimum to $\propto N^{-1}$ for large input photon numbers ($N\ge 10$). The resulting proportionality constants are 1.820 and 1.810 in panels (a) and (b), respectively.
The inset in panel (a) is a zoom for small $N$ values. It shows that odd $N$ input Fock states yield slightly lower phase uncertainties than even $N$ states. The numerical calculations for a coherent pump were carried out up to $N=100$ due to memory limitations.}
\label{f_Min} 
\end{figure}
\end{center}
\vspace{-0.5cm}
\

We next investigate the optimal phase sensitivity achieved once the above-mentioned saturation behaviour has been reached.
To this end, we indicate by vertical lines in Fig.~\ref{f_Unc} the first local minimum of $\Delta\phi$, as well as its lowest minimum in the range of $\tau$ studied here.
Those $\Delta\phi$ minima occur at interaction strengths $\tau$ labeled by $\tau_1$ and $\tau_\mathrm{min}$, respectively.
For a fixed nonlinear coupling strength $\kappa$, the crystal lengths have to be chosen appropriately to obtain these optimal phase sensitivities.
Note that they vary for different input states and different mean number $N$ of input pump photons.

We plot the phase uncertainty $\Delta\phi$ at $\tau_1$ and $\tau_\mathrm{min}$ as a function of $N$ in Fig.~\ref{f_Min} for a pump in either a Fock (top) or a coherent (bottom) state.
In both cases, we observe that the first $\Delta\phi$ minimum is below the shot-noise level $N^{-1/2}$, which is indicated in Fig.~\ref{f_Min} by an orange dotted line.
We also observe for both pump states that the first minimum approaches a Heisenberg scaling ($\Delta\phi(\tau_1)\propto N^{-1}$) for large $N$, as the fit (blue solid line) suggests.
Furthermore, for a pump in a coherent state, the first minimum approaches the shot-noise level for small $N<1$.
This trend is almost inappreciable when the pump is in a Fock state because we are restricted to integer $N$ values, and therefore to $N\geq 1$.
However, we observe a deviation from the Heisenberg scaling for $N$ approaching unity.

For the lowest $\Delta\phi$ minimum in Fig.~\ref{f_Min}, and a coherent pump, the phase uncertainty almost coincides with the first $\Delta\phi$ minimum.
Hence, we also observe a Heisenberg scaling in the lowest minimum for large input numbers $N$, and the uncertainty approaches the shot-noise level for small $N$.
In contrast, for a pump in a Fock state the lowest minimum is noticeably smaller than the first minimum, even though it seems to display a Heisenberg scaling.
For this case, we also observe that the lowest minimum is not a monotonic function of $N$, as highlighted in the inset of Fig.~\ref{f_Min}(a).
In particular, input states for which $N$ is even appear to give slightly worse phase sensitivities.
We present an explanation for this remarkable feature in Sec.~\ref{sec_Inside}.

\section{Photon statistics inside the interferometer}
\label{sec_Inside}

\begin{figure}[htbp]
\centering
\includegraphics[width=\linewidth]{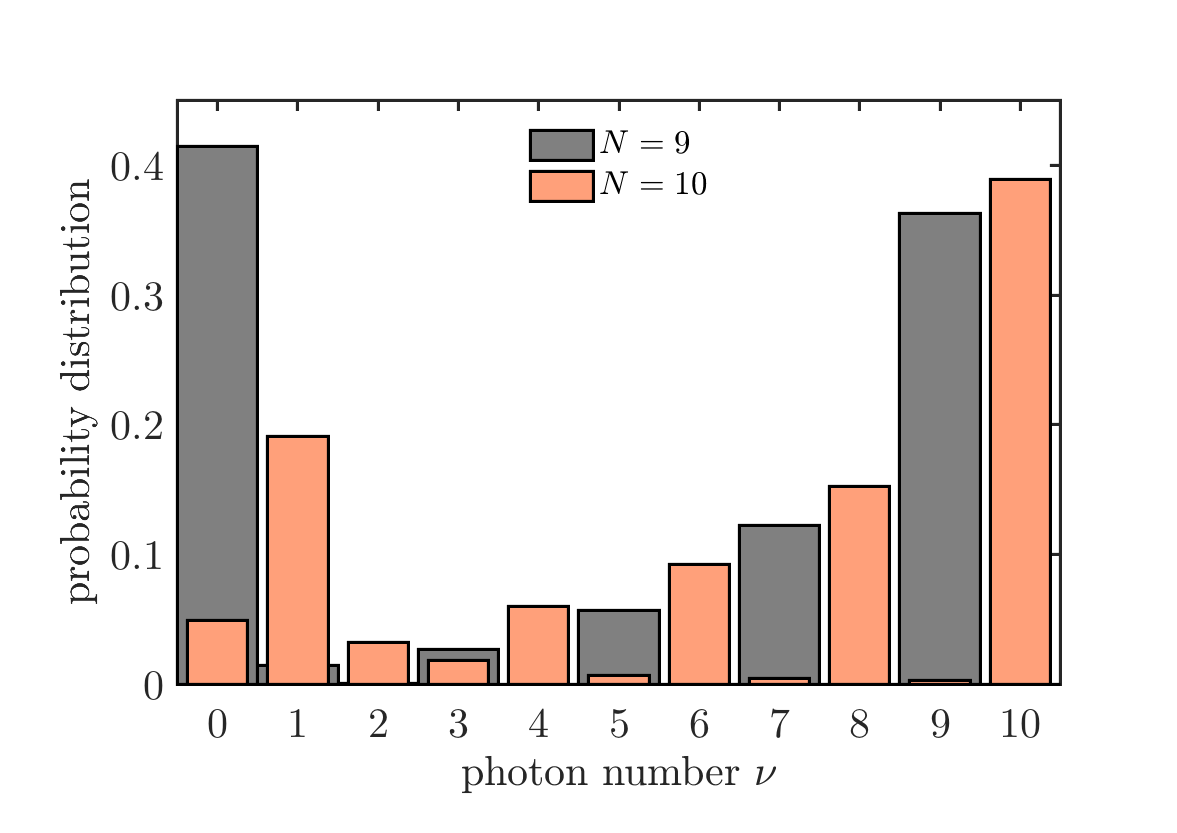}
\caption{Photon number distribution of the signal (or idler) field inside the nonlinear interferometer at maximum phase sensitivity ($\tau=\tau_\mathrm{min}$) for a pump initially in a Fock state.
The mean number of input pump photons are $N=9$ and $10$, and the behaviour is representative for all other two consecutive $N$ values. 
Odd and even input pump photon numbers have a significantly different distribution, which may be the source of their different phase sensitivities in the inset of Fig.~\ref{f_Min}(a).}
\label{f_distribution} 
\end{figure}

To gain more insight into the Heisenberg scaling and the lowest phase uncertainty observed in Sec.~\ref{sec_Uncertainty}, we investigate the quantum state inside the interferometer.
For that, we focus on the simpler case of a pump in a Fock state, and calculate the photon number distribution $|c_\nu|^2$ after crystal $A$ from Eq.~\eqref{e_cA}.
For two exemplary input Fock states ($N=9$ and $10$) and interaction strength $\tau_\text{min}$, we plot $|c_\nu|^2$ in Fig.~\ref{f_distribution}.

For $N=9$, two prominent peaks appear at $\nu=0$ and $\nu=N$.
These two peaks correspond to a superposition of the case where the pump remains in its initial Fock state and where all pump photons are converted to signal and idler field, giving a form similar to $\ket{0}^{(N)}+\ket{N}^{(N)}$.
Using Eq.~\eqref{e_basis}, this state may be written as $\ket{N}_p\ket{0}_s \ket{0}_i+\ket{0}_p\ket{N}_s \ket{N}_i$.
Such a structure resembles a $N00N$ state, $|N0\rangle+|0N\rangle$, in which all $N$ photons appear in either the first or second mode of a linear interferometer~\cite{Sanders89,Dowling02}.
In our case, these two modes are the pump and the signal (and idler) modes. 
$N00N$ states are known to reach the Heisenberg limit with a phase sensitivity $\Delta\phi=1/N$ \cite{Bollinger96,Mitchell04}. 
It is then plausible to assume that this structure leads to the lowest phase uncertainty.

\

Furthermore, in Fig.~\ref{f_distribution} we see oscillations in the photon number distribution.
The distribution goes to zero for even values of $\nu$, as it was previously reported  \cite{Walls70}.
These oscillations arise from destructive interference: Upon time evolution in amplifier $A$, the initial state $\ket{0}^{(N)}$ is depopulated and the population moves towards higher $\nu$.
Since the basis from Eq.~\eqref{e_basis} is finite, the population reflects at $\ket{N}^{(N)}$, resulting in the destructive interference that can be seen in Fig.~\ref{f_distribution}.

When we consider the state $N=10$, we observe a similar structure, but the pronounced two peaks occur at $\nu=N$ and $\nu=1$, rather than $\nu=N$ and $\nu=0$, as one would expect for a $N00N$ state.
This distinction between odd and even $N$ produces the non-monotonic behavior for the lowest $\Delta\phi$ minimum in the inset of Fig.~\ref{f_Min}(a), where the phase sensitivity tends to be better for odd rather than for even $N$.
We then attribute such non-monotonic behavior to the different structures of the quantum states inside the interferometer.
In fact, we find similar photon distributions for all other input intensities:
For odd and even $N$ we always find two peaks, with one at of $\nu=N$. However, for even $N$ the second peak is at $\nu=1$, whereas for odd $N$ it is at $\nu=0$, like a $N00N$ state. 

In contrast, the photon distribution of internal signal photons is almost uniform for an interaction strength $\tau_1$, where the first minimum of the phase uncertainty occurs.
We therefore do not observe a pronounced two-peaked structure that resembles a $N00N$ state.
So, the first minimum of $\Delta \phi$ seems to be of different physical origin.
One can calculate the amount of squeezing of the state inside the interferometer, as in Ref.~\cite{Drobny92}.
In fact, there is a local maximum in the amount of squeezing at a $\tau$ near to $\tau_1$, but they do not coincide exactly.
A local maximum in $N_\mathrm{int}$ is also near to $\tau_1$, but, again, they do not exactly coincide.
It is possibly a combination of a high $N_\mathrm{int}$ and squeezing, rather than a $N00N$-like number distribution, that leads to the first $\Delta \phi$ minimum.

If the pump is initially in a coherent state, the previous analysis can be generalized and it is still possible to observe a two peaked-structure in the joint photon number distribution of pump and signal (and idler) photons inside the interferometer at $\tau=\tau_\mathrm{min}$.
However, in contrast to the Fock state, the distinction between odd or even $N$ seen in Fig.~\ref{f_Min}(b) is absent.
This is roughly what is expected since the coherent state is a superposition of odd and even Fock states and therefore the different distinct features, as described above, wash out.
Indeed, we see that the first and lowest $\Delta\phi$ minimum are of similar order of magnitude in Fig.~\ref{f_Min}(b).

\section{\label{sec_Concl}Conclusions}

We have conducted a rigorous quantum analysis of a nonlinear interferometer, including the quantum nature of the pump field.
In the high-gain regime, where pump depletion and quantum features of the pump are of relevance, the phase uncertainty of an NLI oscillates around a saturation level.
This contrasts with the exponential growth in the low-gain regime described by the parametric approximation.
We further demonstrated that the phase sensitivity is not determined solely by the number of signal (or idler) photons inside the interferometer, but it is also a result of quantum features of the joint state of the pump, signal, and idler fields inside the interferometer.
Most importantly, we showed that the phase uncertainty of the NLI for optimal interaction strengths is below the shot-noise level of a Mach-Zehnder interferometer with the same input intensity.
In fact, the sensitivity of an NLI displays a Heisenberg scaling as the mean number of input pump photons is increased, even when pumped by a coherent state. Finally, we observed that the lowest phase uncertainty occurs when the photon number distribution of the three fields inside the interferometer resembles a $N00N$ state.

A possible extension of our model is the use of two pump beams, one for each amplifier, e.g. created from the output of a beam splitter.
Another extension would be to incorporate dephasing or loss terms. 
Such dephasing and time-ordering effects may become of relevance as well in the high-gain regime~\cite{Christ13}.

In conclusion, we interpret the pump field as the primary resource, rather than the number of photons generated by the parametric amplifiers.
Since in the low-gain regime the number of converted photons is small compared to the input laser intensity conventionally used in a Mach-Zehnder interferometer, the most suitable implementation of the NLI is in the high-gain regime.
Indeed, in this regime we find a Heisenberg scaling and therefore an advantage of the NLI over a conventional Mach-Zehnder interferometer.


\begin{acknowledgments}
This work was supported by the Canada Research Chairs (CRC) Program, the Natural Sciences and Engineering Research Council (NSERC), the Cananda Excellence Research Chairs (CERC) Program, and the Canada First Research Excellence Fund award on Transformative Quantum Technologies.
JF acknowledges support from COLCIENCIAS.
\end{acknowledgments}

\bibliography{References}

\begin{thebibliography}{33}%
\makeatletter
\providecommand \@ifxundefined [1]{%
 \@ifx{#1\undefined}
}%
\providecommand \@ifnum [1]{%
 \ifnum #1\expandafter \@firstoftwo
 \else \expandafter \@secondoftwo
 \fi
}%
\providecommand \@ifx [1]{%
 \ifx #1\expandafter \@firstoftwo
 \else \expandafter \@secondoftwo
 \fi
}%
\providecommand \natexlab [1]{#1}%
\providecommand \enquote  [1]{``#1''}%
\providecommand \bibnamefont  [1]{#1}%
\providecommand \bibfnamefont [1]{#1}%
\providecommand \citenamefont [1]{#1}%
\providecommand \href@noop [0]{\@secondoftwo}%
\providecommand \href [0]{\begingroup \@sanitize@url \@href}%
\providecommand \@href[1]{\@@startlink{#1}\@@href}%
\providecommand \@@href[1]{\endgroup#1\@@endlink}%
\providecommand \@sanitize@url [0]{\catcode `\\12\catcode `\$12\catcode
  `\&12\catcode `\#12\catcode `\^12\catcode `\_12\catcode `\%12\relax}%
\providecommand \@@startlink[1]{}%
\providecommand \@@endlink[0]{}%
\providecommand \url  [0]{\begingroup\@sanitize@url \@url }%
\providecommand \@url [1]{\endgroup\@href {#1}{\urlprefix }}%
\providecommand \urlprefix  [0]{URL }%
\providecommand \Eprint [0]{\href }%
\providecommand \doibase [0]{http://dx.doi.org/}%
\providecommand \selectlanguage [0]{\@gobble}%
\providecommand \bibinfo  [0]{\@secondoftwo}%
\providecommand \bibfield  [0]{\@secondoftwo}%
\providecommand \translation [1]{[#1]}%
\providecommand \BibitemOpen [0]{}%
\providecommand \bibitemStop [0]{}%
\providecommand \bibitemNoStop [0]{.\EOS\space}%
\providecommand \EOS [0]{\spacefactor3000\relax}%
\providecommand \BibitemShut  [1]{\csname bibitem#1\endcsname}%
\let\auto@bib@innerbib\@empty
\bibitem [{\citenamefont {Caves}(1981)}]{Caves81}%
  \BibitemOpen
  \bibfield  {author} {\bibinfo {author} {\bibfnamefont {C.~M.}\ \bibnamefont
  {Caves}},\ }\bibfield  {title} {\enquote {\bibinfo {title}
  {Quantum-mechanical noise in an interferometer},}\ }\href {\doibase
  10.1103/PhysRevD.23.1693} {\bibfield  {journal} {\bibinfo  {journal} {Phys.
  Rev. D}\ }\textbf {\bibinfo {volume} {23}},\ \bibinfo {pages} {1693}
  (\bibinfo {year} {1981})}\BibitemShut {NoStop}%
\bibitem [{\citenamefont {Yurke}\ \emph {et~al.}(1986)\citenamefont {Yurke},
  \citenamefont {McCall},\ and\ \citenamefont {Klauder}}]{Yurke86}%
  \BibitemOpen
  \bibfield  {author} {\bibinfo {author} {\bibfnamefont {B.}~\bibnamefont
  {Yurke}}, \bibinfo {author} {\bibfnamefont {S.~L.}\ \bibnamefont {McCall}}, \
  and\ \bibinfo {author} {\bibfnamefont {J.~R.}\ \bibnamefont {Klauder}},\
  }\bibfield  {title} {\enquote {\bibinfo {title} {{SU}(2) and {SU}(1,1)
  interferometers},}\ }\href {\doibase 10.1103/PhysRevA.33.4033} {\bibfield
  {journal} {\bibinfo  {journal} {Phys. Rev. A}\ }\textbf {\bibinfo {volume}
  {33}},\ \bibinfo {pages} {4033} (\bibinfo {year} {1986})}\BibitemShut
  {NoStop}%
\bibitem [{\citenamefont {Chekhova}\ and\ \citenamefont
  {Ou}(2016)}]{Chekhova16}%
  \BibitemOpen
  \bibfield  {author} {\bibinfo {author} {\bibfnamefont {M.~V.}\ \bibnamefont
  {Chekhova}}\ and\ \bibinfo {author} {\bibfnamefont {Z.~Y.}\ \bibnamefont
  {Ou}},\ }\bibfield  {title} {\enquote {\bibinfo {title} {Nonlinear
  interferometers in quantum optics},}\ }\href {\doibase 10.1364/AOP.8.000104}
  {\bibfield  {journal} {\bibinfo  {journal} {Adv. Opt. Photon.}\ }\textbf
  {\bibinfo {volume} {8}},\ \bibinfo {pages} {104} (\bibinfo {year}
  {2016})}\BibitemShut {NoStop}%
\bibitem [{\citenamefont {Hudelist}\ \emph {et~al.}(2014)\citenamefont
  {Hudelist}, \citenamefont {Kong}, \citenamefont {Liu}, \citenamefont {Jing},
  \citenamefont {Ou},\ and\ \citenamefont {Zhang}}]{Hudelist14}%
  \BibitemOpen
  \bibfield  {author} {\bibinfo {author} {\bibfnamefont {F.}~\bibnamefont
  {Hudelist}}, \bibinfo {author} {\bibfnamefont {J.}~\bibnamefont {Kong}},
  \bibinfo {author} {\bibfnamefont {C.}~\bibnamefont {Liu}}, \bibinfo {author}
  {\bibfnamefont {J.}~\bibnamefont {Jing}}, \bibinfo {author} {\bibfnamefont
  {Z.~Y.}\ \bibnamefont {Ou}}, \ and\ \bibinfo {author} {\bibfnamefont
  {W.}~\bibnamefont {Zhang}},\ }\bibfield  {title} {\enquote {\bibinfo {title}
  {Quantum metrology with parametric amplifier-based photon correlation
  interferometers},}\ }\href {http://dx.doi.org/10.1038/ncomms4049} {\bibfield
  {journal} {\bibinfo  {journal} {Nat. Commun.}\ }\textbf {\bibinfo {volume}
  {5}},\ \bibinfo {pages} {3049} (\bibinfo {year} {2014})}\BibitemShut
  {NoStop}%
\bibitem [{\citenamefont {Kalashnikov}\ \emph {et~al.}(2016)\citenamefont
  {Kalashnikov}, \citenamefont {Paterova}, \citenamefont {Kulik},\ and\
  \citenamefont {Krivitsky}}]{Kalashnikov16}%
  \BibitemOpen
  \bibfield  {author} {\bibinfo {author} {\bibfnamefont {D.~A.}\ \bibnamefont
  {Kalashnikov}}, \bibinfo {author} {\bibfnamefont {A.~V.}\ \bibnamefont
  {Paterova}}, \bibinfo {author} {\bibfnamefont {S.~P.}\ \bibnamefont {Kulik}},
  \ and\ \bibinfo {author} {\bibfnamefont {L.~A.}\ \bibnamefont {Krivitsky}},\
  }\bibfield  {title} {\enquote {\bibinfo {title} {Infrared spectroscopy with
  visible light},}\ }\href {http://dx.doi.org/10.1038/nphoton.2015.252}
  {\bibfield  {journal} {\bibinfo  {journal} {Nat. Photonics}\ }\textbf
  {\bibinfo {volume} {10}},\ \bibinfo {pages} {98} (\bibinfo {year}
  {2016})}\BibitemShut {NoStop}%
\bibitem [{\citenamefont {Barreto~Lemos}\ \emph {et~al.}(2014)\citenamefont
  {Barreto~Lemos}, \citenamefont {Borish}, \citenamefont {Cole}, \citenamefont
  {Ramelow}, \citenamefont {Lapkiewicz},\ and\ \citenamefont
  {Zeilinger}}]{BarretoLemos14}%
  \BibitemOpen
  \bibfield  {author} {\bibinfo {author} {\bibfnamefont {G.}~\bibnamefont
  {Barreto~Lemos}}, \bibinfo {author} {\bibfnamefont {V.}~\bibnamefont
  {Borish}}, \bibinfo {author} {\bibfnamefont {G.~D.}\ \bibnamefont {Cole}},
  \bibinfo {author} {\bibfnamefont {S.}~\bibnamefont {Ramelow}}, \bibinfo
  {author} {\bibfnamefont {R.}~\bibnamefont {Lapkiewicz}}, \ and\ \bibinfo
  {author} {\bibfnamefont {A.}~\bibnamefont {Zeilinger}},\ }\bibfield  {title}
  {\enquote {\bibinfo {title} {Quantum imaging with undetected photons},}\
  }\href {http://dx.doi.org/10.1038/nature13586} {\bibfield  {journal}
  {\bibinfo  {journal} {Nature}\ }\textbf {\bibinfo {volume} {512}},\ \bibinfo
  {pages} {409} (\bibinfo {year} {2014})}\BibitemShut {NoStop}%
\bibitem [{\citenamefont {Lemieux}\ \emph {et~al.}(2016)\citenamefont
  {Lemieux}, \citenamefont {Manceau}, \citenamefont {Sharapova}, \citenamefont
  {Tikhonova}, \citenamefont {Boyd}, \citenamefont {Leuchs},\ and\
  \citenamefont {Chekhova}}]{Lemieux16}%
  \BibitemOpen
  \bibfield  {author} {\bibinfo {author} {\bibfnamefont {S.}~\bibnamefont
  {Lemieux}}, \bibinfo {author} {\bibfnamefont {M.}~\bibnamefont {Manceau}},
  \bibinfo {author} {\bibfnamefont {P.~R.}\ \bibnamefont {Sharapova}}, \bibinfo
  {author} {\bibfnamefont {O.~V.}\ \bibnamefont {Tikhonova}}, \bibinfo {author}
  {\bibfnamefont {R.~W.}\ \bibnamefont {Boyd}}, \bibinfo {author}
  {\bibfnamefont {G.}~\bibnamefont {Leuchs}}, \ and\ \bibinfo {author}
  {\bibfnamefont {M.~V.}\ \bibnamefont {Chekhova}},\ }\bibfield  {title}
  {\enquote {\bibinfo {title} {Engineering the {F}requency {S}pectrum of
  {B}right {S}queezed {V}acuum via {G}roup {V}elocity {D}ispersion in an
  {SU}(1,1) interferometer},}\ }\href {\doibase 10.1103/PhysRevLett.117.183601}
  {\bibfield  {journal} {\bibinfo  {journal} {Phys. Rev. Lett.}\ }\textbf
  {\bibinfo {volume} {117}},\ \bibinfo {pages} {183601} (\bibinfo {year}
  {2016})}\BibitemShut {NoStop}%
\bibitem [{\citenamefont {Linnemann}\ \emph {et~al.}(2016)\citenamefont
  {Linnemann}, \citenamefont {Strobel}, \citenamefont {Muessel}, \citenamefont
  {Schulz}, \citenamefont {Lewis-Swan}, \citenamefont {Kheruntsyan},\ and\
  \citenamefont {Oberthaler}}]{Linnemann16}%
  \BibitemOpen
  \bibfield  {author} {\bibinfo {author} {\bibfnamefont {D.}~\bibnamefont
  {Linnemann}}, \bibinfo {author} {\bibfnamefont {H.}~\bibnamefont {Strobel}},
  \bibinfo {author} {\bibfnamefont {W.}~\bibnamefont {Muessel}}, \bibinfo
  {author} {\bibfnamefont {J.}~\bibnamefont {Schulz}}, \bibinfo {author}
  {\bibfnamefont {R.~J.}\ \bibnamefont {Lewis-Swan}}, \bibinfo {author}
  {\bibfnamefont {K.~V.}\ \bibnamefont {Kheruntsyan}}, \ and\ \bibinfo {author}
  {\bibfnamefont {M.~K.}\ \bibnamefont {Oberthaler}},\ }\bibfield  {title}
  {\enquote {\bibinfo {title} {Quantum-{E}nhanced {S}ensing {B}ased on {T}ime
  {R}eversal of {N}onlinear {D}ynamics},}\ }\href {\doibase
  10.1103/PhysRevLett.117.013001} {\bibfield  {journal} {\bibinfo  {journal}
  {Phys. Rev. Lett.}\ }\textbf {\bibinfo {volume} {117}},\ \bibinfo {pages}
  {013001} (\bibinfo {year} {2016})}\BibitemShut {NoStop}%
\bibitem [{\citenamefont {Chen}\ \emph {et~al.}(2015)\citenamefont {Chen},
  \citenamefont {Qiu}, \citenamefont {Chen}, \citenamefont {Guo}, \citenamefont
  {Chen}, \citenamefont {Ou},\ and\ \citenamefont {Zhang}}]{Chen15}%
  \BibitemOpen
  \bibfield  {author} {\bibinfo {author} {\bibfnamefont {B.}~\bibnamefont
  {Chen}}, \bibinfo {author} {\bibfnamefont {C.}~\bibnamefont {Qiu}}, \bibinfo
  {author} {\bibfnamefont {S.}~\bibnamefont {Chen}}, \bibinfo {author}
  {\bibfnamefont {J.}~\bibnamefont {Guo}}, \bibinfo {author} {\bibfnamefont
  {L.~Q.}\ \bibnamefont {Chen}}, \bibinfo {author} {\bibfnamefont {Z.~Y.}\
  \bibnamefont {Ou}}, \ and\ \bibinfo {author} {\bibfnamefont {W.}~\bibnamefont
  {Zhang}},\ }\bibfield  {title} {\enquote {\bibinfo {title} {Atom-{L}ight
  {H}ybrid {I}nterferometer},}\ }\href {\doibase
  10.1103/PhysRevLett.115.043602} {\bibfield  {journal} {\bibinfo  {journal}
  {Phys. Rev. Lett.}\ }\textbf {\bibinfo {volume} {115}},\ \bibinfo {pages}
  {043602} (\bibinfo {year} {2015})}\BibitemShut {NoStop}%
\bibitem [{\citenamefont {Boyd}(2008)}]{Boyd08}%
  \BibitemOpen
  \bibfield  {author} {\bibinfo {author} {\bibfnamefont {R.~W.}\ \bibnamefont
  {Boyd}},\ }\href@noop {} {\emph {\bibinfo {title} {Nonlinear optics}}}\
  (\bibinfo  {publisher} {Academic press},\ \bibinfo {address} {Cambridge},\
  \bibinfo {year} {2008})\BibitemShut {NoStop}%
\bibitem [{\citenamefont {Manceau}\ \emph
  {et~al.}(2017{\natexlab{a}})\citenamefont {Manceau}, \citenamefont {Leuchs},
  \citenamefont {Khalili},\ and\ \citenamefont {Chekhova}}]{Manceau17b}%
  \BibitemOpen
  \bibfield  {author} {\bibinfo {author} {\bibfnamefont {M.}~\bibnamefont
  {Manceau}}, \bibinfo {author} {\bibfnamefont {G.}~\bibnamefont {Leuchs}},
  \bibinfo {author} {\bibfnamefont {F.}~\bibnamefont {Khalili}}, \ and\
  \bibinfo {author} {\bibfnamefont {M.~V.}\ \bibnamefont {Chekhova}},\
  }\bibfield  {title} {\enquote {\bibinfo {title} {Detection {L}oss {T}olerant
  {S}upersensitive {P}hase {M}easurement with an {SU}(1,1) {I}nterferometer},}\
  }\href {\doibase 10.1103/PhysRevLett.119.223604} {\bibfield  {journal}
  {\bibinfo  {journal} {Phys. Rev. Lett.}\ }\textbf {\bibinfo {volume} {119}},\
  \bibinfo {pages} {223604} (\bibinfo {year} {2017}{\natexlab{a}})}\BibitemShut
  {NoStop}%
\bibitem [{\citenamefont {Plick}\ \emph {et~al.}(2010)\citenamefont {Plick},
  \citenamefont {Dowling},\ and\ \citenamefont {Agarwal}}]{Plick10}%
  \BibitemOpen
  \bibfield  {author} {\bibinfo {author} {\bibfnamefont {W.~N.}\ \bibnamefont
  {Plick}}, \bibinfo {author} {\bibfnamefont {J.~P.}\ \bibnamefont {Dowling}},
  \ and\ \bibinfo {author} {\bibfnamefont {G.~S.}\ \bibnamefont {Agarwal}},\
  }\bibfield  {title} {\enquote {\bibinfo {title} {Coherent-light-boosted,
  sub-shot noise, quantum interferometry},}\ }\href
  {http://stacks.iop.org/1367-2630/12/i=8/a=083014} {\bibfield  {journal}
  {\bibinfo  {journal} {New J. Phys.}\ }\textbf {\bibinfo {volume} {12}},\
  \bibinfo {pages} {083014} (\bibinfo {year} {2010})}\BibitemShut {NoStop}%
\bibitem [{\citenamefont {Li}\ \emph {et~al.}(2014)\citenamefont {Li},
  \citenamefont {Yuan}, \citenamefont {Ou},\ and\ \citenamefont
  {Zhang}}]{Li14}%
  \BibitemOpen
  \bibfield  {author} {\bibinfo {author} {\bibfnamefont {D.}~\bibnamefont
  {Li}}, \bibinfo {author} {\bibfnamefont {C.-H.}\ \bibnamefont {Yuan}},
  \bibinfo {author} {\bibfnamefont {Z.~Y.}\ \bibnamefont {Ou}}, \ and\ \bibinfo
  {author} {\bibfnamefont {W.}~\bibnamefont {Zhang}},\ }\bibfield  {title}
  {\enquote {\bibinfo {title} {The phase sensitivity of an {SU}(1,1)
  interferometer with coherent and squeezed-vacuum light},}\ }\href
  {http://stacks.iop.org/1367-2630/16/i=7/a=073020} {\bibfield  {journal}
  {\bibinfo  {journal} {New J. Phys.}\ }\textbf {\bibinfo {volume} {16}},\
  \bibinfo {pages} {073020} (\bibinfo {year} {2014})}\BibitemShut {NoStop}%
\bibitem [{\citenamefont {Sparaciari}\ \emph {et~al.}(2016)\citenamefont
  {Sparaciari}, \citenamefont {Olivares},\ and\ \citenamefont
  {Paris}}]{Sparaciari16}%
  \BibitemOpen
  \bibfield  {author} {\bibinfo {author} {\bibfnamefont {C.}~\bibnamefont
  {Sparaciari}}, \bibinfo {author} {\bibfnamefont {S.}~\bibnamefont
  {Olivares}}, \ and\ \bibinfo {author} {\bibfnamefont {M.~G.~A.}\ \bibnamefont
  {Paris}},\ }\bibfield  {title} {\enquote {\bibinfo {title} {Gaussian-state
  interferometry with passive and active elements},}\ }\href {\doibase
  10.1103/PhysRevA.93.023810} {\bibfield  {journal} {\bibinfo  {journal} {Phys.
  Rev. A}\ }\textbf {\bibinfo {volume} {93}},\ \bibinfo {pages} {023810}
  (\bibinfo {year} {2016})}\BibitemShut {NoStop}%
\bibitem [{\citenamefont {Mollow}\ and\ \citenamefont
  {Glauber}(1967)}]{Mollow67}%
  \BibitemOpen
  \bibfield  {author} {\bibinfo {author} {\bibfnamefont {B.~R.}\ \bibnamefont
  {Mollow}}\ and\ \bibinfo {author} {\bibfnamefont {R.~J.}\ \bibnamefont
  {Glauber}},\ }\bibfield  {title} {\enquote {\bibinfo {title} {Quantum theory
  of parametric amplification. {I}},}\ }\href {\doibase
  10.1103/PhysRev.160.1076} {\bibfield  {journal} {\bibinfo  {journal} {Phys.
  Rev.}\ }\textbf {\bibinfo {volume} {160}},\ \bibinfo {pages} {1076} (\bibinfo
  {year} {1967})}\BibitemShut {NoStop}%
\bibitem [{\citenamefont {Hamel}\ \emph {et~al.}(2014)\citenamefont {Hamel},
  \citenamefont {Shalm}, \citenamefont {H{\"u}bel}, \citenamefont {Miller},
  \citenamefont {Marsili}, \citenamefont {Verma}, \citenamefont {Mirin},
  \citenamefont {Nam}, \citenamefont {Resch},\ and\ \citenamefont
  {Jennewein}}]{Hamel14}%
  \BibitemOpen
  \bibfield  {author} {\bibinfo {author} {\bibfnamefont {D.~R.}\ \bibnamefont
  {Hamel}}, \bibinfo {author} {\bibfnamefont {L.~K.}\ \bibnamefont {Shalm}},
  \bibinfo {author} {\bibfnamefont {H.}~\bibnamefont {H{\"u}bel}}, \bibinfo
  {author} {\bibfnamefont {A.~J.}\ \bibnamefont {Miller}}, \bibinfo {author}
  {\bibfnamefont {F.}~\bibnamefont {Marsili}}, \bibinfo {author} {\bibfnamefont
  {V.~B.}\ \bibnamefont {Verma}}, \bibinfo {author} {\bibfnamefont {R.~P.}\
  \bibnamefont {Mirin}}, \bibinfo {author} {\bibfnamefont {S.~W.}\ \bibnamefont
  {Nam}}, \bibinfo {author} {\bibfnamefont {K.~J.}\ \bibnamefont {Resch}}, \
  and\ \bibinfo {author} {\bibfnamefont {T.}~\bibnamefont {Jennewein}},\
  }\bibfield  {title} {\enquote {\bibinfo {title} {Direct generation of
  three-photon polarization entanglement},}\ }\href
  {http://dx.doi.org/10.1038/nphoton.2014.218} {\bibfield  {journal} {\bibinfo
  {journal} {Nat. Photonics}\ }\textbf {\bibinfo {volume} {8}},\ \bibinfo
  {pages} {801} (\bibinfo {year} {2014})}\BibitemShut {NoStop}%
\bibitem [{\citenamefont {Ding}\ \emph {et~al.}(2017)\citenamefont {Ding},
  \citenamefont {Maslennikov}, \citenamefont {Habl\"utzel}, \citenamefont
  {Loh},\ and\ \citenamefont {Matsukevich}}]{Ding17}%
  \BibitemOpen
  \bibfield  {author} {\bibinfo {author} {\bibfnamefont {S.}~\bibnamefont
  {Ding}}, \bibinfo {author} {\bibfnamefont {G.}~\bibnamefont {Maslennikov}},
  \bibinfo {author} {\bibfnamefont {R.}~\bibnamefont {Habl\"utzel}}, \bibinfo
  {author} {\bibfnamefont {H.}~\bibnamefont {Loh}}, \ and\ \bibinfo {author}
  {\bibfnamefont {D.}~\bibnamefont {Matsukevich}},\ }\bibfield  {title}
  {\enquote {\bibinfo {title} {Quantum {P}arametric {O}scillator with {T}rapped
  {I}ons},}\ }\href {\doibase 10.1103/PhysRevLett.119.150404} {\bibfield
  {journal} {\bibinfo  {journal} {Phys. Rev. Lett.}\ }\textbf {\bibinfo
  {volume} {119}},\ \bibinfo {pages} {150404} (\bibinfo {year}
  {2017})}\BibitemShut {NoStop}%
\bibitem [{\citenamefont {Drobn\'y}\ \emph {et~al.}(1993)\citenamefont
  {Drobn\'y}, \citenamefont {Jex},\ and\ \citenamefont
  {Bu\ifmmode~\check{z}\else \v{z}\fi{}ek}}]{Drobny93}%
  \BibitemOpen
  \bibfield  {author} {\bibinfo {author} {\bibfnamefont {G.}~\bibnamefont
  {Drobn\'y}}, \bibinfo {author} {\bibfnamefont {I.}~\bibnamefont {Jex}}, \
  and\ \bibinfo {author} {\bibfnamefont {V.}~\bibnamefont
  {Bu\ifmmode~\check{z}\else \v{z}\fi{}ek}},\ }\bibfield  {title} {\enquote
  {\bibinfo {title} {Mode entanglement in nondegenerate down-conversion with
  quantized pump},}\ }\href {\doibase 10.1103/PhysRevA.48.569} {\bibfield
  {journal} {\bibinfo  {journal} {Phys. Rev. A}\ }\textbf {\bibinfo {volume}
  {48}},\ \bibinfo {pages} {569} (\bibinfo {year} {1993})}\BibitemShut
  {NoStop}%
\bibitem [{\citenamefont {Dicke}(1954)}]{Dicke54}%
  \BibitemOpen
  \bibfield  {author} {\bibinfo {author} {\bibfnamefont {R.~H.}\ \bibnamefont
  {Dicke}},\ }\bibfield  {title} {\enquote {\bibinfo {title} {Coherence in
  spontaneous radiation processes},}\ }\href {\doibase 10.1103/PhysRev.93.99}
  {\bibfield  {journal} {\bibinfo  {journal} {Phys. Rev.}\ }\textbf {\bibinfo
  {volume} {93}},\ \bibinfo {pages} {99} (\bibinfo {year} {1954})}\BibitemShut
  {NoStop}%
\bibitem [{\citenamefont {Tavis}\ and\ \citenamefont
  {Cummings}(1968)}]{Tavis68}%
  \BibitemOpen
  \bibfield  {author} {\bibinfo {author} {\bibfnamefont {M.}~\bibnamefont
  {Tavis}}\ and\ \bibinfo {author} {\bibfnamefont {F.~W.}\ \bibnamefont
  {Cummings}},\ }\bibfield  {title} {\enquote {\bibinfo {title} {Exact solution
  for an {$N$}-molecule---radiation-field {H}amiltonian},}\ }\href {\doibase
  10.1103/PhysRev.170.379} {\bibfield  {journal} {\bibinfo  {journal} {Phys.
  Rev.}\ }\textbf {\bibinfo {volume} {170}},\ \bibinfo {pages} {379} (\bibinfo
  {year} {1968})}\BibitemShut {NoStop}%
\bibitem [{\citenamefont {Tucker}\ and\ \citenamefont
  {Walls}(1969)}]{Tucker69}%
  \BibitemOpen
  \bibfield  {author} {\bibinfo {author} {\bibfnamefont {J.}~\bibnamefont
  {Tucker}}\ and\ \bibinfo {author} {\bibfnamefont {D.~F.}\ \bibnamefont
  {Walls}},\ }\bibfield  {title} {\enquote {\bibinfo {title} {Quantum theory of
  the traveling-wave frequency converter},}\ }\href {\doibase
  10.1103/PhysRev.178.2036} {\bibfield  {journal} {\bibinfo  {journal} {Phys.
  Rev.}\ }\textbf {\bibinfo {volume} {178}},\ \bibinfo {pages} {2036} (\bibinfo
  {year} {1969})}\BibitemShut {NoStop}%
\bibitem [{\citenamefont {Walls}\ and\ \citenamefont
  {Barakat}(1970)}]{Walls70}%
  \BibitemOpen
  \bibfield  {author} {\bibinfo {author} {\bibfnamefont {D.~F.}\ \bibnamefont
  {Walls}}\ and\ \bibinfo {author} {\bibfnamefont {R.}~\bibnamefont
  {Barakat}},\ }\bibfield  {title} {\enquote {\bibinfo {title}
  {Quantum-mechanical amplification and frequency conversion with a trilinear
  {H}amiltonian},}\ }\href {\doibase 10.1103/PhysRevA.1.446} {\bibfield
  {journal} {\bibinfo  {journal} {Phys. Rev. A}\ }\textbf {\bibinfo {volume}
  {1}},\ \bibinfo {pages} {446} (\bibinfo {year} {1970})}\BibitemShut {NoStop}%
\bibitem [{\citenamefont {Bonifacio}\ and\ \citenamefont
  {Preparata}(1970)}]{Bonifacio70}%
  \BibitemOpen
  \bibfield  {author} {\bibinfo {author} {\bibfnamefont {R.}~\bibnamefont
  {Bonifacio}}\ and\ \bibinfo {author} {\bibfnamefont {G.}~\bibnamefont
  {Preparata}},\ }\bibfield  {title} {\enquote {\bibinfo {title} {Coherent
  spontaneous emission},}\ }\href {\doibase 10.1103/PhysRevA.2.336} {\bibfield
  {journal} {\bibinfo  {journal} {Phys. Rev. A}\ }\textbf {\bibinfo {volume}
  {2}},\ \bibinfo {pages} {336} (\bibinfo {year} {1970})}\BibitemShut {NoStop}%
\bibitem [{\citenamefont {Drobn\'y}\ and\ \citenamefont
  {Jex}(1992)}]{Drobny92}%
  \BibitemOpen
  \bibfield  {author} {\bibinfo {author} {\bibfnamefont {G.}~\bibnamefont
  {Drobn\'y}}\ and\ \bibinfo {author} {\bibfnamefont {I.}~\bibnamefont {Jex}},\
  }\bibfield  {title} {\enquote {\bibinfo {title} {Quantum properties of field
  modes in trilinear optical processes},}\ }\href {\doibase
  10.1103/PhysRevA.46.499} {\bibfield  {journal} {\bibinfo  {journal} {Phys.
  Rev. A}\ }\textbf {\bibinfo {volume} {46}},\ \bibinfo {pages} {499} (\bibinfo
  {year} {1992})}\BibitemShut {NoStop}%
\bibitem [{\citenamefont {Manceau}\ \emph
  {et~al.}(2017{\natexlab{b}})\citenamefont {Manceau}, \citenamefont
  {Khalili},\ and\ \citenamefont {Chekhova}}]{Manceau17a}%
  \BibitemOpen
  \bibfield  {author} {\bibinfo {author} {\bibfnamefont {M.}~\bibnamefont
  {Manceau}}, \bibinfo {author} {\bibfnamefont {F.}~\bibnamefont {Khalili}}, \
  and\ \bibinfo {author} {\bibfnamefont {M.~V.}\ \bibnamefont {Chekhova}},\
  }\bibfield  {title} {\enquote {\bibinfo {title} {Improving the phase
  super-sensitivity of squeezing-assisted interferometers by squeeze factor
  unbalancing},}\ }\href {http://stacks.iop.org/1367-2630/19/i=1/a=013014}
  {\bibfield  {journal} {\bibinfo  {journal} {New J. Phys.}\ }\textbf {\bibinfo
  {volume} {19}},\ \bibinfo {pages} {013014} (\bibinfo {year}
  {2017}{\natexlab{b}})}\BibitemShut {NoStop}%
\bibitem [{\citenamefont {Giese}\ \emph {et~al.}(2017)\citenamefont {Giese},
  \citenamefont {Lemieux}, \citenamefont {Manceau}, \citenamefont {Fickler},\
  and\ \citenamefont {Boyd}}]{Giese17}%
  \BibitemOpen
  \bibfield  {author} {\bibinfo {author} {\bibfnamefont {E.}~\bibnamefont
  {Giese}}, \bibinfo {author} {\bibfnamefont {S.}~\bibnamefont {Lemieux}},
  \bibinfo {author} {\bibfnamefont {M.}~\bibnamefont {Manceau}}, \bibinfo
  {author} {\bibfnamefont {R.}~\bibnamefont {Fickler}}, \ and\ \bibinfo
  {author} {\bibfnamefont {R.~W.}\ \bibnamefont {Boyd}},\ }\bibfield  {title}
  {\enquote {\bibinfo {title} {Phase sensitivity of gain-unbalanced nonlinear
  interferometers},}\ }\href {\doibase 10.1103/PhysRevA.96.053863} {\bibfield
  {journal} {\bibinfo  {journal} {Phys. Rev. A}\ }\textbf {\bibinfo {volume}
  {96}},\ \bibinfo {pages} {053863} (\bibinfo {year} {2017})}\BibitemShut
  {NoStop}%
\bibitem [{\citenamefont {Marino}\ \emph {et~al.}(2012)\citenamefont {Marino},
  \citenamefont {Corzo~Trejo},\ and\ \citenamefont {Lett}}]{Marino12}%
  \BibitemOpen
  \bibfield  {author} {\bibinfo {author} {\bibfnamefont {A.~M.}\ \bibnamefont
  {Marino}}, \bibinfo {author} {\bibfnamefont {N.~V.}\ \bibnamefont
  {Corzo~Trejo}}, \ and\ \bibinfo {author} {\bibfnamefont {P.~D.}\ \bibnamefont
  {Lett}},\ }\bibfield  {title} {\enquote {\bibinfo {title} {Effect of losses
  on the performance of an {SU}(1,1) interferometer},}\ }\href {\doibase
  10.1103/PhysRevA.86.023844} {\bibfield  {journal} {\bibinfo  {journal} {Phys.
  Rev. A}\ }\textbf {\bibinfo {volume} {86}},\ \bibinfo {pages} {023844}
  (\bibinfo {year} {2012})}\BibitemShut {NoStop}%
\bibitem [{\citenamefont {Gerry}\ and\ \citenamefont {Knight}(2004)}]{Gerry04}%
  \BibitemOpen
  \bibfield  {author} {\bibinfo {author} {\bibfnamefont {C.}~\bibnamefont
  {Gerry}}\ and\ \bibinfo {author} {\bibfnamefont {P.}~\bibnamefont {Knight}},\
  }\enquote {\bibinfo {title} {Beam splitters and interferometers},}\ in\ \href
  {\doibase 10.1017/CBO9780511791239.006} {\emph {\bibinfo {booktitle}
  {Introductory Quantum Optics}}}\ (\bibinfo  {publisher} {Cambridge University
  Press},\ \bibinfo {year} {2004})\ p.\ \bibinfo {pages} {135}\BibitemShut
  {NoStop}%
\bibitem [{\citenamefont {Sanders}(1989)}]{Sanders89}%
  \BibitemOpen
  \bibfield  {author} {\bibinfo {author} {\bibfnamefont {B.~C.}\ \bibnamefont
  {Sanders}},\ }\bibfield  {title} {\enquote {\bibinfo {title} {Quantum
  dynamics of the nonlinear rotator and the effects of continual spin
  measurement},}\ }\href {\doibase 10.1103/PhysRevA.40.2417} {\bibfield
  {journal} {\bibinfo  {journal} {Phys. Rev. A}\ }\textbf {\bibinfo {volume}
  {40}},\ \bibinfo {pages} {2417--2427} (\bibinfo {year} {1989})}\BibitemShut
  {NoStop}%
\bibitem [{\citenamefont {Lee}\ \emph {et~al.}(2002)\citenamefont {Lee},
  \citenamefont {Kok},\ and\ \citenamefont {Dowling}}]{Dowling02}%
  \BibitemOpen
  \bibfield  {author} {\bibinfo {author} {\bibfnamefont {H.}~\bibnamefont
  {Lee}}, \bibinfo {author} {\bibfnamefont {P.}~\bibnamefont {Kok}}, \ and\
  \bibinfo {author} {\bibfnamefont {J.~P.}\ \bibnamefont {Dowling}},\
  }\bibfield  {title} {\enquote {\bibinfo {title} {A quantum rosetta stone for
  interferometry},}\ }\href {\doibase 10.1080/0950034021000011536} {\bibfield
  {journal} {\bibinfo  {journal} {Journal of Modern Optics}\ }\textbf {\bibinfo
  {volume} {49}},\ \bibinfo {pages} {2325--2338} (\bibinfo {year} {2002})},\
  \Eprint {http://arxiv.org/abs/https://doi.org/10.1080/0950034021000011536}
  {https://doi.org/10.1080/0950034021000011536} \BibitemShut {NoStop}%
\bibitem [{\citenamefont {Bollinger}\ \emph {et~al.}(1996)\citenamefont
  {Bollinger}, \citenamefont {Itano}, \citenamefont {Wineland},\ and\
  \citenamefont {Heinzen}}]{Bollinger96}%
  \BibitemOpen
  \bibfield  {author} {\bibinfo {author} {\bibfnamefont {J.~J~.}\ \bibnamefont
  {Bollinger}}, \bibinfo {author} {\bibfnamefont {W.~M.}\ \bibnamefont
  {Itano}}, \bibinfo {author} {\bibfnamefont {D.~J.}\ \bibnamefont {Wineland}},
  \ and\ \bibinfo {author} {\bibfnamefont {D.~J.}\ \bibnamefont {Heinzen}},\
  }\bibfield  {title} {\enquote {\bibinfo {title} {Optimal frequency
  measurements with maximally correlated states},}\ }\href {\doibase
  10.1103/PhysRevA.54.R4649} {\bibfield  {journal} {\bibinfo  {journal} {Phys.
  Rev. A}\ }\textbf {\bibinfo {volume} {54}},\ \bibinfo {pages} {R4649}
  (\bibinfo {year} {1996})}\BibitemShut {NoStop}%
\bibitem [{\citenamefont {Mitchell}\ \emph {et~al.}(2004)\citenamefont
  {Mitchell}, \citenamefont {Lundeen},\ and\ \citenamefont
  {Steinberg}}]{Mitchell04}%
  \BibitemOpen
  \bibfield  {author} {\bibinfo {author} {\bibfnamefont {M.~W.}\ \bibnamefont
  {Mitchell}}, \bibinfo {author} {\bibfnamefont {J.~S.}\ \bibnamefont
  {Lundeen}}, \ and\ \bibinfo {author} {\bibfnamefont {A.~M.}\ \bibnamefont
  {Steinberg}},\ }\bibfield  {title} {\enquote {\bibinfo {title}
  {Super-resolving phase measurements with a multiphoton entangled state},}\
  }\href {http://dx.doi.org/10.1038/nature02493} {\bibfield  {journal}
  {\bibinfo  {journal} {Nature}\ }\textbf {\bibinfo {volume} {429}},\ \bibinfo
  {pages} {161} (\bibinfo {year} {2004})}\BibitemShut {NoStop}%
\bibitem [{\citenamefont {Christ}\ \emph {et~al.}(2013)\citenamefont {Christ},
  \citenamefont {Brecht}, \citenamefont {Mauerer},\ and\ \citenamefont
  {Silberhorn}}]{Christ13}%
  \BibitemOpen
  \bibfield  {author} {\bibinfo {author} {\bibfnamefont {A.}~\bibnamefont
  {Christ}}, \bibinfo {author} {\bibfnamefont {B.}~\bibnamefont {Brecht}},
  \bibinfo {author} {\bibfnamefont {W.}~\bibnamefont {Mauerer}}, \ and\
  \bibinfo {author} {\bibfnamefont {C.}~\bibnamefont {Silberhorn}},\ }\bibfield
   {title} {\enquote {\bibinfo {title} {Theory of quantum frequency conversion
  and type-{II} parametric down-conversion in the high-gain regime},}\ }\href
  {http://stacks.iop.org/1367-2630/15/i=5/a=053038} {\bibfield  {journal}
  {\bibinfo  {journal} {New J. Phys.}\ }\textbf {\bibinfo {volume} {15}},\
  \bibinfo {pages} {053038} (\bibinfo {year} {2013})}\BibitemShut {NoStop}%
\end{thebibliography}%

\appendix

\section{\label{app_CFI}Fisher information}
In Sec.~\ref{sec_Uncertainty}, we pointed out that the phase can be obtained from the mean number $N_\mathrm{out}$ of signal (or idler) photons at the output of the NLI.
Thus, we estimated the phase uncertainty from error propagation of $N_\mathrm{out}$, Eq.~\eqref{e_EP}. 
However, we may use any other estimator formula for the phase based on the output signal (or idler, or pump) statistics, like the root mean squared of signal photons, just to mention one example. 
In this Appendix, we investigate the best phase sensitivity that can be reached based on the output signal photon statistics.
This phase sensitivity is provided by the Fisher information,
\begin{equation}
\mathcal{F} = \sum_{N_\mathrm{out}} \frac{1}{P(N_\mathrm{out}|\phi)}\left(\frac{\partial P(N_\mathrm{out}|\phi)}{\partial \phi}\right)^2,
\label{e_FC}
\end{equation}
with $P(N_\mathrm{out}|\phi)$ being the probability of measuring $N_\mathrm{out}$ output signal photons given a certain phase $\phi$. 
This probability is calculated from Eq.~\eqref{e_cB}.
The phase uncertainty $\Delta\phi_\mathrm{FI}$ from the Fisher information is then given by $1/\sqrt{\mathcal{F}}$.

\begin{figure}[htbp]
\centering
\includegraphics[width=\linewidth]{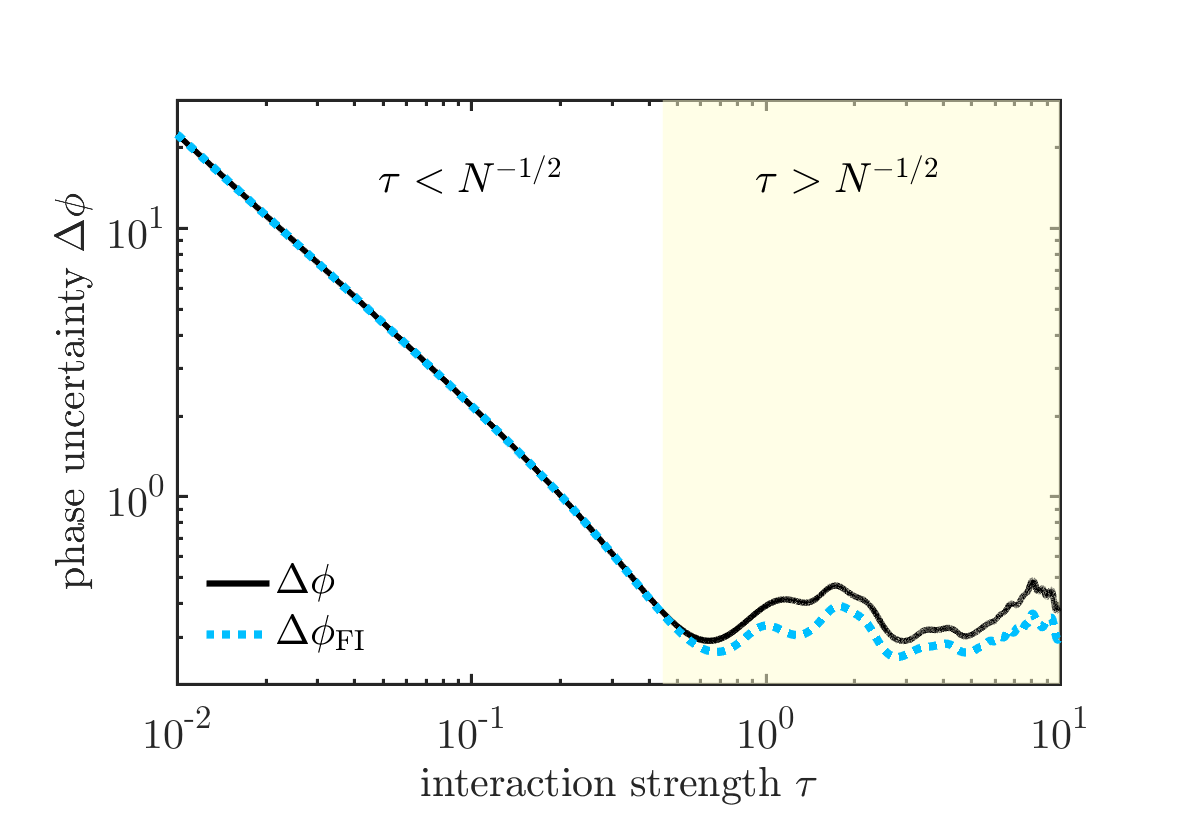}
\caption{Comparison of the phase uncertainty estimated by error propagation, Eq.~\eqref{e_EP}, and the classical Fisher information, Eq.~\eqref{e_FC}, of the output signal (or pump, or idler) photon number distribution. We chose a coherent input pump field with a mean number of photons $N=5$. The yellow shadow area defines the high-gain regime, $\tau>N^{-1/2}$.}
\label{f_UncFC}
\end{figure}
\

According to the Cram\'er-Rao bound, the phase uncertainty $\Delta\phi_\mathrm{FI}$ limits the phase uncertainty from below, i.e. $\Delta\phi_\mathrm{FI}\le\Delta\phi$, with $\Delta\phi$ given by Eq.~\eqref{e_EP}. 
We emphasize that, even though we know that the phase uncertainty is bounded by the Fisher information, the estimator itself is not specified.
In contrast, for error propagation, we are simply using the mean number of output signal photons as an estimator. 
In Fig.~\ref{f_UncFC}, we compare the results for $\Delta\phi_\mathrm{FI}$ as a function of the interaction strength $\tau$ to the ones obtained from error propagation.

On one hand, we observe that in the low-gain regime, $\tau<N^{-1/2}$, Eq.~\eqref{e_EP} and Eq.~\eqref{e_FC} lead to the same phase uncertainty.
On the other hand, in the high-gain regime, $\tau>N^{-1/2}$, the general trend of $\Delta\phi$ and $\Delta\phi_\mathrm{FI}$ is approximately the same, although the uncertainty obtained from the Fisher information is slightly smaller, as expected from the Cram\'er-Rao bound. 
Moreover, $\tau_1$ and $\tau_\mathrm{min}$ for $\Delta\phi$ and $\Delta\phi_\mathrm{CF}$ are very close to each other, but do not exactly coincide.

\begin{figure}[h!]
\centering
\includegraphics[width=\linewidth]{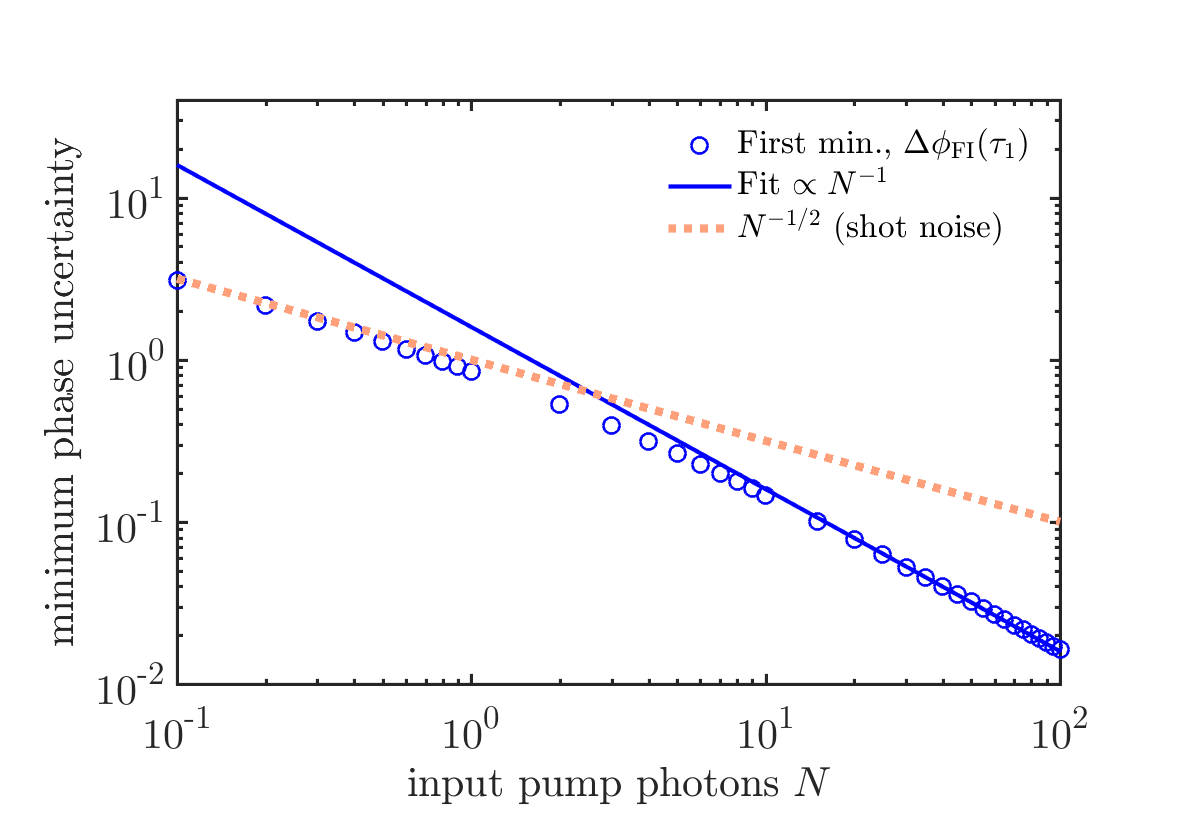}
\caption{First minimum of the phase uncertainty $\Delta\phi_\mathrm{FI}$ as a function of the mean number $N$ of input pump photons for a coherent pump.
We show a Heisenberg scaling by fitting the first minimum of $\Delta\phi_\mathrm{FI}$ to $\propto N^{-1}$ for $N\ge 10$. The resulting proportionality constant is 1.591.}
\label{f_MinFC} 
\end{figure}
\

To investigate the influence of the slightly reduced Fisher information phase uncertainty, we follow the procedure from Sec.~\ref{sec_Uncertainty}, and show in Fig.~\ref{f_MinFC} the first minimum of $\Delta\phi_\mathrm{FI}$ as a function of the mean number $N$ of input pump photons.
We again observe a phase uncertainty that approaches the shot-noise level (orange dotted line) from below for small $N$ ($N<1$).
For large $N$, we observe a Heisenberg scaling highlighted by a fit (blue solid line) in Fig.~\ref{f_MinFC}. Likewise, for the lowest $\Delta\phi_\mathrm{FI}$ minimum over the range investigated, and for the pump in a Fock state, we observe qualitatively the same results as the uncertainties discussed in the main part of the article, even though they are slightly smaller.
However, since the overall behaviour is the same, we refrain from presenting these results for brevity.
\end{document}